\documentclass[aps,amsmath,prb,twocolumn,showpacs]{revtex4}
\usepackage{epsf}

\newcommand{\beq}{\begin{equation}}
\newcommand{\eeq}{\end{equation}}
\newcommand{\beqa}{\begin{eqnarray}}
\newcommand{\eeqa}{\end{eqnarray}}
\newcommand{\w}{\omega}
\newcommand{\W}{\Omega}
\newcommand{\e}{\epsilon}
\newcommand{\nn}{{\bf n}}
\renewcommand{\r}{{\bf r}}
\renewcommand{\P}{{\bf P}}
\newcommand{\0}{{\bf 0}}
\newcommand{\R}{{\bf R}}
\renewcommand{\k}{{\bf k}}
\newcommand{\q}{{\bf q}}
\newcommand{\p}{{\bf p}}
\newcommand{\Q}{{\bf Q}}
\newcommand{\sgn}{\mbox{sgn}}

\renewcommand{\Im}{{\rm Im}}

\begin{document}           

\title{Two dimensional electron gas near full polarization} 

\author{G. Zala}
\affiliation{Department of Physics \& Astronomy, Stony Brook University, \\ 
Stony Brook, NY 11794, USA}
\author{B. N. Narozhny}
\affiliation{Condensed Matter Section, ICTP, \\
Strada Costiera 11, I-34100, Trieste, Italy}
\author{I. L. Aleiner}
\affiliation{Physics Department, Columbia University, \\ 
538 West 120th Street, 
New York, NY 10027, USA} 
\author{Vladimir I. Fal'ko}
\affiliation{Physics Department, Lancaster University, \\ 
LA1 4YB, Lancaster, UK} 

\date{\today}

\begin{abstract}             

We establish the consistency of the Fermi liquid description and find
a relation between Fermi liquid constants for the two dimensional
electron system near the point of full polarization due to a parallel
magnetic field $H$.  Our results enable us to predict connections
between different thermodynamic properties of the system.  In
particular, we find that near the point of full polarization $H_c$,
the thermodynamic compressibility of the system experiences a jump
with the subleading $(H_c-H)^{1/2}$ dependence on the magnetic field.
Also, the magnetization has a cusp with the dependence of the type
$(H-H_c) + (H_c-H)^{3/2}$ at $H<H_c$.
\end{abstract} 

\maketitle   

\section{Introduction}
\label{strongfield}

The appearance of the new generation of high mobility heterostructures
\cite{pfeiffer} resulted in observations of interesting phenomena
\cite{effects,du,sivan} which brought new spotlight on the effects of
strong interactions in two-dimensional electron systems (2DES). For us
this renews theoretical interest in the thermodynamical properties of
clean Fermi liquids.

Traditional theoretical description of 2DES is based upon the theories
of weakly interacting Fermi gas \cite{ando} and Landau Fermi liquid
theory \cite{landau,AGD}. In the limit of very strong interactions
(corresponding to low electron densities) such a description breaks
down. However the situation changes if the low density 2DES 
interacts with another 2DES with much higher density. In this
paper we present an example where this coexistance naturally occurs as
a result of a strong spin polarization of a 2DES due to a high
in-plane magnetic field. We find a consistent Landau Fermi liquid
description for this system (despite the fact that a naive estimate of
the plasma parameter $r_s$ for the minority spin component yields a
formally large value). The remarkable feature of our result is that in
the close vicinity of the spin polarized state the perturbative
expansion is possible in terms of the inverse gas parameter of the low
density subsystem. This enbles us to determine the functional form of
the dependencies of the 2DES compressibility, magnetization, and
specific heat on the small density of the minority electrons.

To understand the relation of this problem to the Fermi liquid theory
at zero field let us recall the basic structure of the quasiparticle
interaction functional (we will not write the trivial long-range
Coulomb interaction term)

\beq
{\cal H}_{\rm int}=\frac{1}{2}F^\rho \rho^2 + \frac{1}{2}
\sum_{\alpha,\beta=x,y,z}F^{\sigma}_{\alpha\beta}S^{\alpha}S^{\beta},
\label{zerofield}
\eeq 

\noindent
where $\rho$ is the charge density, $S^{x,y,z}$ denote components of
the spin density, and $F^\rho, \ \hat{F}^{\sigma}$ are the
corresponding Fermi liquid parameters. The charge (singlet channel)
and the spin (triplet) fluctuations are decoupled, and $SU(2)$
symmetry of the system guarantees
$F^{\sigma}_{\alpha\beta}=F^{\sigma}\delta_{\alpha\beta}$.

If the magnetic field is applied along, say, the $x$ - direction (in
the plane of the 2DES), the $SU(2)$ symmetry is reduced to $U(1)$ and one
may write for the quadratic part of the energy

 \beq
{\cal H}_{\rm int}=\frac{F^\rho}{2} \rho^2 +
\frac{F^{\sigma}_{\parallel}}{2}\left[S^{x}\right]^2
+ {F^{\sigma\rho}} \rho S^{x}
+ \frac{F^{\sigma}_{\perp}}{2}
\sum_{\alpha=y,z}S^{\alpha}S^{\alpha}.
\label{1field}
\eeq

\noindent
This means that the system can no longer be described by two constants. 
Now the reduced symmetry allows for four independent parameters.

Simplifications are possible, however, with the further increase of
the magnetic field, because oscillations of the spin density
components $S_{y,z}$ become gapped (the gap equals to the Zeeman
splitting $E_z$). Therefore, for the description of low lying
excitations with the energy much smaller than $E_z$, the last term in
Eq.~(\ref{1field}) can be ignored. Introducing deviations of densities
of majority (minority) electrons $n_{1(2)}=\rho/2 \pm S_x/2$, one
obtains the two-fluid model

\beq
{\cal H}_{\rm int}=\frac{F_{11}}{2} n_1^2 +
\frac{F_{22}}{2} n_2^2 + F_{12}n_1n_2
\label{2field}
\eeq

\noindent
characterized by three independent parameters.

Equation (\ref{2field}) suggests two questions: (i) what is the lowest
density of minority electrons for which it is applicable; and (ii)
whether the three constants of the model are indeed independent. The
ultimate goal of this paper is to show that (i) the Fermi liquid
description is consistent for any density of the minority electrons
and (ii) there is a relation between the Fermi liquid constants for
the vanishing density of the minority electrons. The only requirement
for this description to be valid is that the fully polarized electron
system is a stable Fermi liquid. The remarkable feature of this result
is that in the close vicinity of the spin polarized state the
perturbative expansion in terms of the inverse gas parameter $\hbar
v_{F2}/e^2$ is possible (here $v_{F2}$ is the Fermi velocity of the
minority electrons).

The remainder of the paper is organized as follows.  In
Section~\ref{phenom} we give a phenomenological description of the
system near full polarization, present the main results and predict
connections between different thermodynamic properties of the system.
Section~\ref{microder} contains the microscopic derivation
(justification) of the announced results, first in an intuitive, then
in Subsections~\ref{3b} and \ref{3c}, in a more rigorous manner.

\section{Phenomenology near full polarization}
\label{phenom}

\subsection{Structure of the theory}

Let us consider the system first at zero temperature.  Because the
total spin of the system commutes with the Hamiltonian, we can write
the energy density of the system $\delta {\cal H}$ in terms of the
majority ($n_1$) and minority electron density ($n_2$).  Omitting the
trivial term of the direct Coulomb interaction (we will work with
fixed total density $n_1+n_2=N$), one finds

\beqa \label{energy}
&&{\cal H} =
\frac{1+F_{11}}{2\nu_1}  \delta n_1^2 + {\cal H}_2(n_2,N) 
+ \frac{F_{12}}{{\nu_1}}  n_2 \delta n_1 -
\nonumber\\
&&
\nonumber\\
&& \quad\quad
- \frac{E_z - E_z^c(N)}{2} 
\left(\delta n_1- n_2\right) - \frac{E_z}{2} N. 
\eeqa

\noindent
The first term in Eq.~(\ref{energy}) is the quadratic expansion of the
ground state energy of the fully spin polarized electron system,
$n_1=N$. It has the standard Fermi-liquid form with $\nu_1$ being the
density of states (entering the slope of the specific heat). The
second term ${\cal H}_2$ characterizes the energy of the minority
electrons at fixed $n_1=N$ and the third term characterizes the change
in this energy due to modification of the majority density. The last
two terms characterize the shift of the energies due to the magnetic
field, $H$, and $E_z=g\mu_BH$ is the bare Zeeman splitting, with $g$
being the bare (non-renormalized by electron-electron interaction)
Lande $g$-factor, and $\mu_B$ being the Bohr magneton. The quantity
$E_z^c(N)$ corresponds to the value of the magnetic field above which
the magnetization is independent of the field. In other words, this
value limits from above the region of the field where the finite
density of the minority electrons is still energetically profitable.

In order to find the ground state of the whole system we have to
minimize energy Eq.~(\ref{energy}) with respect to the electron
densities. Having in mind that the total electron density is fixed by
an external gate, we note that the densities are coupled by the
constraint

\beq
 \delta n_1 + n_2 = \delta N, \quad n_2\geq 0,
\label{constr}
\eeq

\noindent
where $\delta N$ is the change in the total electron density with
respect to the density threshold for the population of the minority
subband controllable by the variation of the gate voltage. This yields
either $n_2=0$ or

\beqa
\label{der1}
&&0=\frac{\partial {\cal H}_2(n_2,N)}{\partial n_2}+ \left[E_z -
E_z^c(N)\right]+ 
\\
&& 
\nonumber\\
&& \quad\quad
+ \frac{1+F_{11} }{\nu_1} \left(n_2-\delta N\right)-
\frac{F_{12}}{\nu_1}\left(2 n_2-\delta N\right).
\nonumber 
\eeqa 

\noindent
The crititical field $E_z^c(N+\delta N)$ is determined as the field at
which $n_2=0$ solves Eq.~(\ref{der1}). Then the first term vanishes (see
below) and we obtain

\beq
\frac{\partial E_z^c}{\partial N} = 
\frac{1+ F_{11}-F_{12}}{\nu_1}. 
\label{Ec}
\eeq

Further progress requires knowledge of the function ${\cal
H}_2(n_2,N)$. We find

\begin{subequations}
\label{H2}
\beq
\label{H22}
{\cal  H}_2(n_2,N)=
\int\limits_0^{n_2}d n^\prime\int\limits_0^{n^{\prime}}d n^{\prime\prime}
\left[ \frac{1}{\nu_2(n^{\prime\prime},N)}+
\frac{F_{22}(n^{\prime\prime},N)}{\nu_1(N)}\right],
\eeq

\noindent
where the dependence of the density of states of the electron density
is given by

\beqa
&&\frac{1}{\nu_2(n,N)} = 
\frac{1}{\nu_2(N)} + 
\label{nu2}
\\
&& 
\nonumber\\
&& \quad\quad
+ \frac{4\sqrt{\pi n}[1+F_{11}(N)-F_{12}(N)]^2}
{3e^2\pi^2\left[\nu_1(N)\right]^2} + 
{\cal O}\left(\frac{n}{N}\ln\frac{n}{N}\right), 
\nonumber
\eeqa

\noindent
and the Fermi liquid constant for the minority electrons is

\beqa
&& F_{22}(n,N) = -  \left[1+ F_{11} (N)-  2 F_{12}(N) \right] +
\label{myF22}\\
&& 
\nonumber\\
&&\quad +
\frac{4\sqrt{\pi n}[1+F_{11}(N)-F_{12}(N)]^2}{e^2\pi^2\nu_1(N)}
+{\cal O}\left(\frac{n}{N}\ln\frac{n}{N} \right).
\nonumber
\eeqa
\end{subequations}

Equations (\ref{H2}) constitute the key point of this paper. They
state, that properties of the minority electrons can be expressed in
terms of the density of states $\nu_2(N) \equiv \nu_2(n_2=0,N)$ at the
point of full polarization (renormalized by interaction with the
majority electrons) and the Fermi liquid constants, $F_{11},\ F_{12}$
of the majority electrons.  We will see below that the above equations
impose certain connections between different observable quantities. It
is interesting to notice that the relevant expansion parameter here is
not the strength of the Coulomb interaction, $e^2$, but rather its
inverse power. This expansion is valid for $n_2 a_B^2 \ll 1$, where
$a_B$ is the usual screening radius in two dimensions,
$a_B\simeq1/\nu_1 e^2$.

Postponing a rigorous derivation of Eqs.~(\ref{H2}) until the next
section, we discuss their physical meaning. Consider $n_2 \to 0$ and
retain only the first line in Eq.~(\ref{myF22}). Substituting the
result into Eqs.~(\ref{H22}) and (\ref{energy}) and keeping only
quadratic terms we find

\[
{\cal H} = \frac{n_2^2}{2\nu_2}+
\frac{\left(1+F_{11}\right)}{2\nu_1} \left(\delta n_1^2- n_2^2\right) 
 + \frac{F_{12} }{{\nu_1}}n_2\left(n_2 + \delta n_1\right). 
\]

\noindent
Let us now set $\delta n_1 = -n_2$ (keeping electrical
neutrality). The Hamiltonian takes the form

\[
{\cal H} = \frac{n_2^2}{2\nu_2}, 
\]

\noindent
which correponds to the compressibility of {\em non - interacting}
minority electrons. This is not accidental; the majority electrons
screen the Coulomb interaction at distances of the order of the
screening radius. At small densities, however, the distance between
minority electrons is much larger than this radius. Therefore, this
screened interaction is seen by minority electrons only as contact
interaction, i.e. the effect of this interaction vanishes because of
the Pauli principle. This explains the origin of the first line in
Eq.~(\ref{myF22}). The second term in Eq.~(\ref{myF22}) describes the
effect of the finite interaction range. This effect clearly vanishes
as the distance between minority electrons increases.  Because the
residual interaction is of the dipole type $U_{sc}(r)\simeq
e^2a_B^2/r^3$, its effect can be estimated as $\delta {\cal H} \simeq
U_{sc}(r=1/\sqrt{n_2})n_2$ which immediately gives a $\sqrt{n_2}$
dependence to the Fermi liquid parameter $F_{22}$.  As a matter of
fact, the same $\sqrt{n_2}$ arises in all angular harmonics of
$F_{22}$ (and $F_{12}$ as well; see the following section). This leads
to renormalization of the effective mass and, therefore, the density
of states in Eq.~(\ref{nu2}). The residual interaction is weak and
therefore the perturbative treatment of the minority electrons is
legitimate.

To complete the calculation of the ground state energy, we use
Eqs.~(\ref{H2}) in Eq.~(\ref{der1}) and find with the help of
Eq.~(\ref{Ec})

\beqa \label{n2}
&& n_2 = n_0\left[ 1 - \gamma R_s n_0^{1/2} +\dots\right];
\\
&&
\nonumber\\
&&
R_s = \frac{ 1   }{  e^2\nu_2} 
\left(\nu_2\frac{\partial E_z^c}{\partial N}\right)^2 ;
\nonumber\\
&&
\nonumber\\
&&
n_0 =   \nu_2(N) \left[E_z^c(N+\delta N) - 
E_z\right]\theta\left(E_z^c(N+\delta
N) - E_z\right)
\nonumber 
\eeqa

\noindent
Here $\theta(x)$ is the step function, and

\beq
\gamma = \frac{32}{9\pi^{3/2}} = 0.639 \dots
\label{gamma}
\eeq

\noindent
is a numerical coefficient.  It is noteworthy that the subleading term
in this dependence is singular near $E_z=E_z^c$.  Substituting
Eq.~(\ref{n2}) into Eqs.~(\ref{H2}) and the result into
Eq.~(\ref{energy}), we find for the ground state energy ${\cal E}$

\beq
\label{finalenergy}
{\cal E}=
\frac{1+F_{11}}{2\nu_1}\delta N^2
- \frac{E_z N}{2}
- \frac{n_0^2 }{2\nu_2} + \frac{2}{5} \frac{\gamma R_s n_0^{5/2}}{\nu_2},
\eeq

\noindent
where we omitted terms linear in $\delta N$ (apart from the term
proportional to $E_z$) as they will not contribute to the observable
quantities discussed in the following subsection.

\subsection{Experimental consequences.}
\label{experiment}

In this subsection we apply the above ideas to establish relations
between different thermodynamic properties of the system near the
point of full polarization.

The leading contribution to the specific heat of the two-liquid system
is simply the sum of the quasiparticle specific heats of the two
species of electrons,

\beqa
&&\left(\frac{\partial C_V}{\partial T}\right)_{T\to 0} = 
\frac{\pi^2\nu_1(N)}{3} +
\\
&&
\nonumber\\
&&
\quad\quad
+
\frac{\pi^2 \nu_2(N)}{3}\left[1-\left(\frac{3\gamma^2 R_s^2 \nu_2 \delta
    E_z}{8}\right)^{\frac{1}{2}}\right]
\theta\left(\delta E_z\right) ;
\nonumber\\
&&
\nonumber\\
&&\quad\quad\quad\quad\quad
\delta E_z= E_z^c(N)- E_z.
\nonumber
\eeqa
This gives the operational definition of the density of states, even
though the actual measurement of the specific heat in two dimensions
is technically difficult.

Next, Eq.~(\ref{Ec}) allows one to find a certain combination of the
Fermi-liquid constants $1+ F_{11}-F_{12}$ from the measurement of the 
critical magnetic field as a function of electron density.

Further information about the Fermi liquid constants can be obtained
from studying the thermodynamic compressibility $\kappa=(\partial^2
{\cal E}/\partial^2 N )^{-1}$, where ${\cal E}$ is the ground state
energy of the system. Differentiating Eq.~(\ref{finalenergy}) yields

\beq
\frac{1}{\kappa} =
\frac{1+F_{11}}{\nu_1}- e^2 R_s 
\left(1-\frac{3\gamma}{2}
\sqrt{R_s^2 \nu_2 \delta E_z}
\right)
\theta\left(\delta E_z\right).
\label{kappa}
\eeq

\noindent
We see that the measurement of the jump in the compressibility gives
the value of the parameter $R_s$ which together with Eq.~(\ref{n2})
and measurements of $\partial_N E_z^c(N)$ yields the value of
$\nu_2(N)$ without specific heat measurements. After that the
coefficient in front of subleading square-root singularity does not
contain any fitting parameters.

Finally, we calculate the magnetizetion $M=-\partial_H{\cal E}$ as the
function of magnetic field $H$. Differentiating
Eq.~(\ref{finalenergy}) at $\delta N =0$ with the help of
Eq.~(\ref{n2}), we obtain

\beq
M=g\mu_B
\left[\frac{N}{2}-\nu_2\delta E_z
\left(1-\gamma
\sqrt{R_s^2 \nu_2 \delta E_z}
\right)\theta\left(\delta E_z\right)
\right].
\label{M}
\eeq

It is important to emphasize that after the compressibility and $M$ at
$E_z > E_z^c$ are measured, the formula for magnetization at $E_z <
E_z^c$ will not have any adjustable parameters. It is also worth
noticing that the sub-leading dependence has a square root
singularity, similar to that in Eq.~(\ref{kappa}).

The predicted dependencies of magnetization and compressibility are
plotted on Fig.~\ref{observables}.

{ 
\begin{figure}[ht]  
%\vspace{0.1 cm} 
\epsfxsize=8.5 cm  
\centerline{\epsfbox{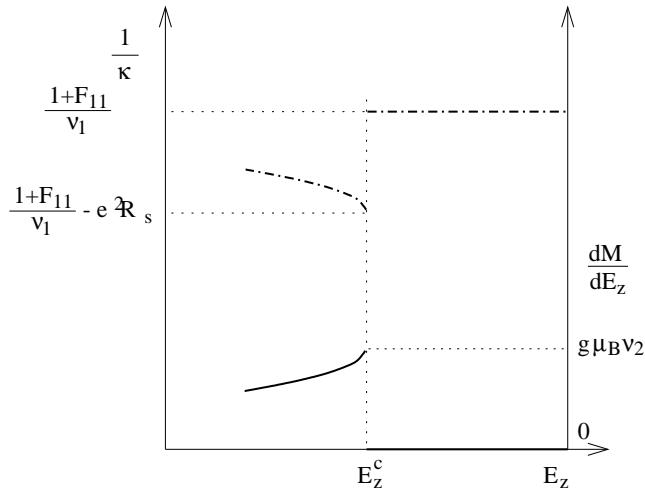}} 
%\vspace{0.2cm} 
\caption{Predicted dependences of compressibility (dashed-dotted line) and 
susceptibility (solid line).}
\label{observables} 
\end{figure} 
}

\section{Microscopic derivation} 
\label{microder}

The purpose of this section is to develop the microscopic description
leading to Eqs.~(\ref{H2}), which is needed to justify the Fermi
liquid description of the minority electrons and to calculate the
coefficient in the second term of Eqs.~(\ref{nu2}) and (\ref{myF22}).
The form of the first term in Eq.~(\ref{myF22}) follows already from
the physical argument presented after Eq.~(\ref{myF22}) and it will
also be confirmed by the microscopic calculation.

The route we are taking in this section is the following. First, we
express our physical arguments in the language of the perturbation
theory, i.e. identify the set of diagrams leading to
Eqs.~(\ref{H2}). The rest of the derivation aims at showing that (i)
these are the only diagrams that produce a combination of the constant
($n_2=0$ term) and the term with $\sqrt{n_2}$ singularity, (ii) all
other diagrams result in contributions of the order of $(n_2/N) \ln
n_2/N $; (iii) the Fermi liquid description of the minirity electrons
is justified.  This material will be structured into subsections
\ref{3a}, \ref{3b} and \ref{3c}. The rather cumbersome content of
these subsections will not directly contribute to the final results
and might be skipped by a pragmatic reader.

\subsection{Perturbation theory}
\label{3a}

The goal of the microscopic consideration presented in this section is
to prove the main assumption of the phenomenological treatment, namely
Eqs.~(\ref{H2}). We start by showing how one can arrive at
Eqs.~(\ref{H2}) using a simple-minded perturbative approach. The
reason to do this is to build up physical intuition, identify the
group of diagrams which gives the dominant contribution to the
minority electron interaction, and to clarify the assumptions, which
one needs to make in order to justify this treatment. In the following
two subsections we shall prove the validity of these assumptions and
provide a more rigorous treatment of the problem.

The main physical idea of the following picture is that close to the
point of full polarization only the majority electrons participate in
screening. Therefore, as a building block for the perturbation theory
we are going to use the dynamically screened (by the majority
electrons) Coulomb interaction, which propagator has a diagrammatic
representation of Fig.~{\ref{screening}}. The corresponding analytic
expression is

\beq
\label{Vqw}
V(\w,q) = \frac{V_0(q)}{1+V_0(q)\Pi_1\left(\omega,q\right)},
\eeq

\noindent
where $V_0(q)=2\pi e^2/|q|$ is the bare Coulomb interaction and
$\Pi_1\left(\omega,q\right)$ is the polarization operator of the
majority system. It is defined as the part of the density - density
correlation function irreducible with respect to one Coulomb line. For
small momentum and frequency transfers, $q,\w/v_{F1} \ll p_{F1}$, it
has the usual Fermi-liquid form \cite{AGD}

\beqa 
&&\Pi_1\left(\omega,q\right)=\int \frac{d\phi_1d\phi_2}{(2\pi)^2}
\Pi_1\left(\omega,\q;{\nn}_1,{\nn}_2\right);
\label{PiFermi}\\
&&
\nonumber\\
&&\Pi_1\left(\omega,\q;{\nn}_1,{\nn}_2\right)=
\Pi_1^{qp}\left(\omega,\q;{\nn}_1\right)
2\pi\delta(\phi_1-\phi_2)-
\nonumber\\ 
&&
\nonumber\\
&& 
- \int
\frac{d\phi_3}{2\pi}
\Pi_1^{qp}\left(\omega,\q;{\nn}_1\right)
\frac{F_{11}\left(\widehat{\nn_1\nn_3}\right)}{\nu_1}
\Pi_1\left(\omega,\q;{\nn}_3,{\nn}_2\right); 
\nonumber\\
&&
\nonumber\\
&&
\Pi_1^{qp}\left(\omega,\q;{\nn}\right)= \frac{\nu_1
v_{F1}{\nn} \cdot \q} {v_{F1}{\nn} \cdot \q
-\w-i0\sgn\w}. 
\label{Piqp}
\eeqa 

\noindent
Here $F_{11}\left(\widehat{\nn_1\nn_2}\right)$ is the majority Fermi
liquid parameter (taken at zero minority density $n_2=0$),
$\Pi_1^{qp}\left(\omega,q;{\nn}\right)$ characterizes the linear
response of non-interacting quasiparticles, and $\nu_1$ is the density
of states of majority quasiparticles. The latter quantity enters into
the specific heat of the spin polarized ($n_2=0$) system. Equation
(\ref{PiFermi}) takes into account all possible contributions singular
as a function of $v_{F1}q/\w$ and neglects all the contributions which
are regular functions of the parameters $(q/p_{F1})^2$ and
$\omega/\e_{F1}$.  Unit vectors $\nn_i=(\cos\phi_i,\sin\phi_i)$
characterize the direction of motion of the quasiparticle.

Now we discuss the origin of Eqs.~(\ref{H2}).

{ 
\begin{figure}[ht]  
%\vspace{0.1 cm} 
\epsfxsize=8 cm  
\centerline{\epsfbox{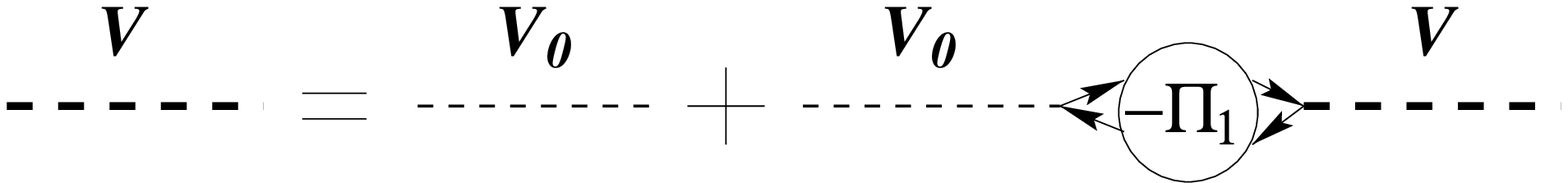}} 
%\vspace{0.2cm} 
\caption{Dyson equation for the dynamical screening by majority electrons.}
\label{screening} 
\end{figure} 
}

\subsubsection{Interaction of minority electrons}

To justify the relation Eq.~(\ref{myF22}) we need to describe the
interaction of minority electrons in terms of the parameters of
majority electrons. Since interaction between minority electrons is
characterized by the energy transfer $\w \simeq qv_{F2} \ll qv_{F1}$,
we can use the static approximation for $V(\w,q)$. Moreover, at
wavectors smaller than the inverse screening radius of majority
electrons $\nu_1V_0(q) \gg 1$. In this case Eq.~(\ref{Vqw}) becomes
(see subsection \ref{3b} for further discussions)

\beq
V\left(\w,q\right) \approx \frac{1}{\nu_1} \left[ 1+F_{11}^{(0)} -
\frac{\left(1+F_{11}^{(0)}\right)^2}{\nu_1 V_0(q)} + \dots \right].
\label{v0}
\eeq

\noindent
Here we use angular harmonics of the Fermi liquid functions

\beq
F_{ij}\left(\phi\right) = \sum_{m}e^{im\phi}F_{ij}^{(m)},
\label{Fm}
\eeq

\noindent
for $i,j=1,2$. The zeroth angular harmonics, $F_{ij}^{(0)}$, that
appear in Eq.~(\ref{v0}) correspond to the constants used in the
previous sections.

{ 
\begin{figure}[ht]  
%\vspace{0.1 cm} 
\epsfxsize=2 cm  
\centerline{\epsfbox{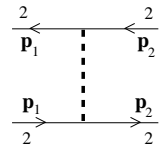}} 
\vspace{0.2cm} 
\caption{Leading contribution to $F_{22}$.}
\label{F22first} 
\end{figure} 
} 

The lowest order contribution to the minority electron interaction
(that determines the Fermi liquid constant $F_{22}$) is given by the
diagram in Fig.~\ref{F22first}. Using Eq.~(\ref{v0}) to evaluate this
contribution we find 

\beq {\rm Fig}.~\ref{F22first}.= -
\frac{1+F_{11}^{(0)}}{\nu_1} +
\frac{\left(1+F_{11}^{(0)}\right)^2}{\nu_1^2 V_0(|\p_1-\p_2|)} \dots.
\label{a1}
\eeq 

\noindent
This contribution is proportional to $|\p_1-\p_2|\sim
p_{F2}\propto\sqrt{n_2}$ which produces the singular density
dependence in Eq.~(\ref{myF22}).

We now show that higher order diagrams built out of the same
ingredients as the simplest diagram in Fig.~\ref{F22first} depend on
at least the second power of $|\p_1-\p_2|$ and thus result only in
regular contributions to $F_{22}$.

Indeed, summation of the ten second order diagrams in
Fig.~\ref{F22second} yields zero whenever one of the dashed lines is
substituted with a constant.  Therefore, the constant part of the
potential Eq.~(\ref{v0}) can be omitted entirely from the second order
perturbation theory and one obtains

\beqa
&&{\rm Fig}.~\ref{F22second}. \ {\rm (a)} + \dots + {\rm (g)} \propto
\left[\frac{\left(1+F_{11}^{(0)}\right)^2}{\nu_1^2
    V_0(p_{12})}\right]^2
\propto (p_{12})^2;
\nonumber \\
&& \quad \quad
\quad p_{12}=|\p_1-\p_2|.
\nonumber
\eeqa

\noindent
This cancellation is not accidental and in fact is due to the Fermi
statistics of the minority electrons. All higher order terms are
canceled in the same manner.

{ 
\begin{figure}[ht]  
%\vspace{0.1 cm} 
\epsfxsize=7 cm  
\centerline{\epsfbox{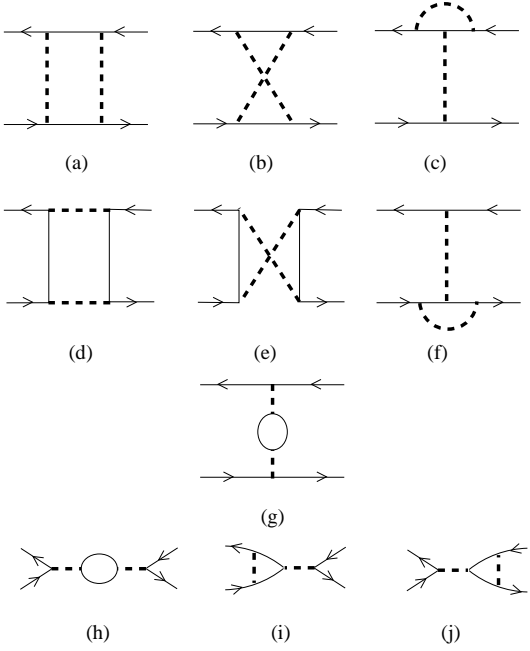}} 
\vspace{0.2cm} 
\caption{Connected second-order minority diagrams (all solid lines
denote Green's functions of minority electrons).}
\label{F22second} 
\end{figure} 
} 

Unfortunately, this is still not the whole story. Majority electrons
affect minority electrons not only through the density-density
interaction but also through the renormalization of the spectrum of
minority electrons (the simplest analogy here is the polaronic shift
of the bottom of the minority band). This renormalization depends on
the distribution function of majority electrons and, therefore,
generates the Fermi liquid function $F_{12}$.  The lowest order
diagrams for this parameter are shown in Fig.~\ref{F12lowest}.
Precisely the same diagrams enter into the two particle irreducible
vertex (that contributes to $F_{22}$) in Fig.~\ref{alles}.  Therefore,
this so far neglected contribution to the minority interaction can be
expressed entirely in terms of $F_{12}$ (by means of direct comparison
with diagrams in Fig.~\ref{F12lowest}). We find

\begin{subequations}
\beqa
&&{\rm Fig}.~\ref{alles}.(a) =  
\frac{[F_{12}^{(0)}]^2}{\nu_1^2}\Pi_1(0, p_{12}) = 
\frac{[F_{12}^{(0)}]^2}{\nu_1 (1+F_{11}^{(0)})}  \\
&&{\rm Fig}.~\ref{alles}.(b) = - 
\frac{[F_{12}^{(0)}]^2}{\nu_1^2}\Pi_1(0,p_{12})^2V(0,
p_{12})
\\
&&
\nonumber\\
&&
\hspace{1.55cm}
=- \frac{[F_{12}^{(0)}]^2}{\nu_1 (1+F_{11}^{(0)})} 
+ \frac{[F_{12}^{(0)}]^2}{\nu_1^2 V_0(p_{12})} 
\nonumber\\
&&
\nonumber\\
&&
{\rm Fig}.~\ref{alles}.(c)= {\rm Fig}.~\ref{alles}.(d) = 
F_{12}^{(0)}\Pi_1(0,p_{12})V(0,p_{12})
\nonumber\\
&&
\nonumber\\
&&
\hspace{1.55cm} = 
\frac{F_{12}^{(0)}}{\nu_1}
\left[1-\frac{1+F_{11}^{(0)}}{\nu_1 V_0(p_{12})}\right]
\eeqa
\label{figs}
\end{subequations}

\noindent
The reason that only the zeroth harmonics of the Fermi liquid
parameters appear in Eq.~(\ref{figs}) is that we assume: (i)
$F_{12}^{(0)}\approx const + {\cal O}(q^2/p_{F1}^2)$ (where $q$ is the
transmitted wavevector); (ii) $F_{12}^{(1)}\sim {\cal
O}(q/p_{F1})$. The dependence of $q$ on the scale of the order of
$p_{F1}$ can be neglected because it would generate smallness of the
order of $n_2/N$.

Finally, to obtain the Fermi liquid parameter $F_{22}$ we combine the 
two-particle irreducible minority vertex functions discussed above:

\[
\frac{F_{22}(\theta)}{\nu_1} = {\rm Fig}.~\ref{F22first} + {\rm
  Fig}.~\ref{alles}(a) + \dots + (d)\Big|_{
\begin{matrix} |\p_i|=p_{F2};\cr \widehat{\p_1\p_2} =\theta, \end{matrix}
}
\]

\noindent
where $i=1,2$.  Summing up contributions of Eq.~(\ref{figs}) and
(\ref{a1}) we obtain the angle-dependent Fermi liquid parameter
$F_{22}$

\beq
F_{22}(\theta) = -  \left[1+ F_{11}^{(0)} -  2 F_{12}^{(0)} \right]+
\frac{[1+F_{11}^{(0)}-F_{12}^{(0)}]^2}{\nu_1 V_0(2p_{F2}\sin\frac{\theta}{2})}.
\label{F22ang}
\eeq

\noindent
The zeroth angular harmonic of Eq.~(\ref{F22ang}) gives precisely
Eq.~(\ref{myF22}), if we recall the Landau theorem that relation
$n_2=p_{F2}^2/4\pi$ is not changed by interaction in any order of the
perturbation theory.

{ 
\begin{figure}[ht]  
%\vspace{0.1 cm} 
\epsfxsize=5 cm  
\centerline{\epsfbox{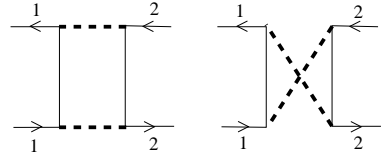}} 
\vspace{0.2cm} 
\caption{Lowest order diagrams for $F_{12}$.}
\label{F12lowest} 
\end{figure} 
} 

\subsubsection{Density of states of minority electrons}

In order to determine the DoS (or the effective mass) of minority
quasiparticles, let us recall Galilean invariance, which results in
the following two-liquid variant of the usual Ward identity

\beq
\frac{1}{m_2(n)} = \frac{1}{m} - \frac{1}{m_1} 
\int \frac{d\theta}{2\pi} 
\left\{ \frac{p_{F1}}{p_{F2}} F_{12}(\theta) + F_{22}(\theta) \right\}
\cos(\theta),
\label{m2gen}
\eeq

\noindent
where $m$ is the bare (band) electron mass, and $m_1$ is the majority
quasiparticle mass.  At zero minority density $F_{12}$ renormalizes
the mass by the amount of order one [since the large factor
$p_{F1}/p_{F2}$ cancels exactly due to the assumptions that we made
deriving Eqs.~(\ref{figs}); see the assumption (ii) in the paragraph
followingEqs.~(\ref{figs})]:

\beq
\frac{1}{m_2(0)} = \frac{1}{m} - \frac{1}{m_1} 
\lim_{p_{F2}\to 0}  \frac{p_{F1}}{p_{F2}} \int \frac{d\theta}{2\pi}  
F_{12}(\theta) 
\cos(\theta).
\label{m20}
\eeq

\noindent
After the cancellation of the prefactor the remaining dependence of
$F_{12}^{(1)}$ on $n_2$ is analytic. Hence, this term does not produce
any singular dependence of the minority DoS on $n_2$. The square root
dependence is caused entirely by the interaction between minority
electrons, i.e. $F_{22}$. By plugging in Eq.~(\ref{F22ang}) to
Eq.~(\ref{m2gen}) and using Eq.~(\ref{m20}) one immediately retrieves
the DoS Eq.~(\ref{nu2}).

{ 
\begin{figure}[ht]  
%\vspace{0.1 cm} 
\epsfxsize=7 cm  
\centerline{\epsfbox{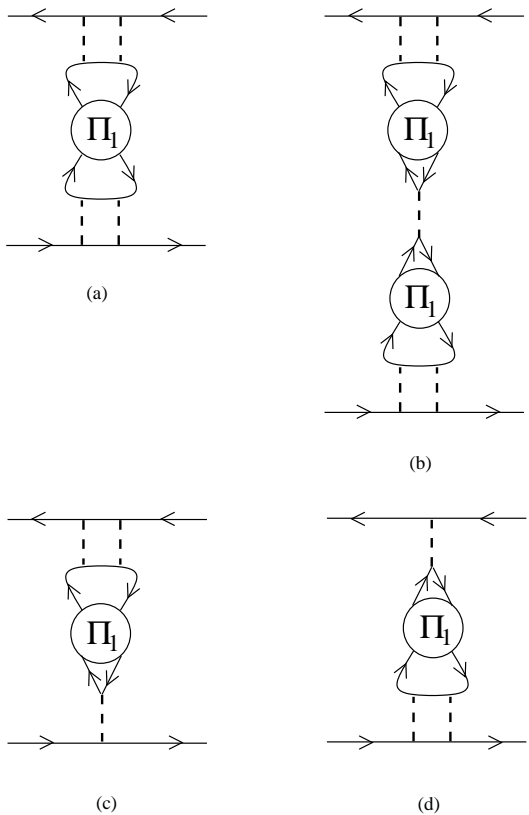}} 
\vspace{0.2cm} 
\caption{Contributions to $F_{22}$ from the renormalization of the
minority spectrum.}
\label{alles} 
\end{figure} 
}

\subsubsection{Underlying assumptions}

The line of argument presented so far relies on several assumptions
which require further justification. These include: (i) the momentum
dependence of $F_{12}$ was assumed to have the specific form (see the
text following Eqs.~(\ref{figs}); (ii) the screened interaction
$V(\w,\q)$ was only considered in the limit of small frequencies,
based on the intuitive assumption $\omega\simeq v_{F2}q \ll v_{F1}q$
[see text preceding Eq.~(\ref{v0})]; (iii) the consideration of
minority interaction was limited to certain class of diagrams, see
above. 

The assumption (i) allowed for the explicit result Eq.~(\ref{F22ang})
that followed from the evaluation of the diagrams in
Figs.~\ref{F22first} and \ref{alles}. The essence of the assumption
(iii) is that no other diagram contributes to the singular dependence
of $F_{22}$. Partially this was illustrated by considering diagrams in
Fig.~\ref{F22second}, however, one could imagine more complicated
diagrams involving majority electrons. Moreover, the cancellation of
diagrams in Fig.~\ref{F22second} relied heavily on the assumption
(ii). Thus, in order to rigorously prove our conjecture
Eqs.~(\ref{H2}) we need to justify the above assumptions. Although
proper consideration of these issues will not change the final
results, we include the following subsections in order to complete the
derivation.

Our strategy will be the following.  First, we will set the number of
minority electrons $n_2$ to zero {\em inside} diagrams in
consideration, and discuss the analytic and scaling properties of the
self-energy and $n$-point irreducible vertex functions.  We will see
that apart from a well-defined subclass of minority 2-particle
vertices, (providing us with the nonanalytic dependence for $F_{22}$),
all n-point functions are smooth (i.e. Taylor-expandable) as a
function of external momenta.  Second, we will treat $n_2 \ll N$ as a
perturbation and show that this may introduce only corrections linear
in the small parameter $n_2/N$.  We will use the vertex functions and
the gauge invariance of the theory to calculate the Fermi liquid
constant $F_{22}$ in terms of the Fermi-liquid parameters of majority
spin and obtain Eq.~(\ref{F22ang}). Finally, we will justify the
calculation of the minority mass in more detail.

\subsection{Completely polarized system}
\label{3b}

In this subsection we discuss the properties of the irreducible vertex
functions which we shall use in the following subsection to calculate
the Fermi liquid parameters.

\subsubsection{Green functions}

We start by defining zero-temperature, real time Green functions of
minority ($j=2$) or majority ($j=1$) electrons: 

\beqa 
&&\langle T_t
\hat{\psi}_j({\bf R}_1)\hat{\psi}_k^\dagger({\bf R}_2) \rangle 
\\
&&
\nonumber\\
&&
\quad=i\delta_{jk }\int \frac{d^3P}{(2\pi)^3} e^{-i{\bf P}({\bf
R}_1-{\bf R}_2)} G_j({\bf P}), 
\nonumber 
\eeqa 

\noindent
where all the operators are taken in the Heisenberg representation and
averaging is performed over the ground state of the system. To shorten
the notation we use hereinafter the $(1+2)$-dimensional vectors ${\bf
R}\equiv (t,\r)$ and ${\bf P}\equiv (\e,\p)$ with the scalar product
${\bf P}{\bf R}=\e t - \p\r$. There are no Green functions mixing the
electron species because the electron spin component along the
magnetic field is conserved.

Let us set the number of minority electrons $n_2$ to zero.  For
majority electrons we will need only the linearized spectrum near the
energy shell:

\beq
\label{gf1}
G_1(\epsilon, {\bf p}) = 
\frac{Z_1}{\epsilon -\xi_1 -
\Sigma_1\left(\displaystyle\frac{\epsilon}{v_{F1}p_{F1}}, 
\displaystyle\frac{\xi_1}{v_{F1}p_{F1}}\right)},
\eeq

\noindent
where $\xi_1=v_{F1}\left(|{\bf p}|-p_{F1}\right)$ is the distance from
the Fermi surface, $Z_1$ is the quasiparticle weight, $v_{F1}$ is the
Fermi velocity renormalized by interaction, and $p_{F1}$ is the Fermi
momentum. The relation $[p_{F1}]^2=4\pi n_1$ ($n_1$ being the density
of majority electrons) is not affected by interaction due to the
conservation of the number of states and the spin conservation (Landau
theorem).  The remainder of the self-energy for the majority electrons
possesses the following property:

\beq
\Sigma_1(x, y) = {\rm Re}\Sigma_1(x, y) 
+ i \sgn x\ \left|{\rm Im}\Sigma_1(x, y)\right|.
\eeq

\noindent
The leading dependence of the self-energy in two dimesnions
is $\Sigma_1(x, x) \simeq i  |x| x \ln(1/|x|)$. 

As usual in Fermi liquid theory, the leading non-analytic
dependences of vertex functions orginate from the overlap of poles
of two Green functions with close momenta. In this case we will use the
standard representation

\beqa
&&G_1\left(\P +\frac{\Q}{2}\right) G_1
\left(\P-\frac{\Q}{2}\right) 
\label{twopoles} \\
&&
\nonumber\\
&&\quad\quad 
= 2i\pi Z_1^2 \delta(\epsilon)\delta(\xi) 
\frac{\Pi_1^{qp}\left(\omega,q;{\nn}\right)}{\nu_1} + 
\varphi(\P,\Q),  \nonumber
\eeqa

\noindent
where 2+1 vector $\Q=(\omega,\ \q)$ is small in comparison with the
Fermi momentum, $\nn=\p/|p|$ is the unit vector along the momentum
$\p$, and $\varphi(\P,\Q)$ is a smooth function with well defined
limit at $\Q\to 0$. The quasiparticle polarization operator was
defined in Eq.~(\ref{Piqp}).

The minority electrons are described in a similar manner with the
exception that now the spectrum cannot be linearized:

\beq
G_2(\epsilon, {\bf p}) = \frac{1}{\epsilon - \frac{p^2}{2m} - 
\Sigma_2(\epsilon, {\bf p})},
\label{g2}
\eeq

\noindent
where $m$ is the bare mass of the electron and $\Sigma_2$ is the
self-energy of minority electrons. Because $n_2 =0$, condition $\int
d\e e^{i0\e }G_2(\epsilon, {\bf p}) =0$ must be satisfied and
therefore $G_2(\epsilon)$ is an analytic function of $\epsilon$ at
${\rm Im}\epsilon >0$.  The self-energy has the effect of
renormalizing the residue, the mass and the chemical potential:

\beqa
- \Sigma_2(\e, p) =&&i0 + Z_2^{-1} \mu_2 + (Z_2^{-1}-1)\e  
\label{sigma2}
\\ 
&& 
\nonumber\\
&&
- \left( \frac{1}{Z_2 m_2} - \frac{1}{m} \right) \frac{p^2}{2} -
Z_2^{-1} \tilde{\Sigma}_2.  
\nonumber 
\eeqa 

\noindent
Here the parameter $Z_2$ is the quasiparticle weight for the minority
electrons. It has the physical meaning of the overlap of the initial
wave-function of the majority electrons with the wave-function of
these electrons after they screen the potential of an introduced
minority electron. The chemical potential $\mu_2$ is shifted with
respect to its bare zero value by the interaction with the majority
electrons. This effect is analogous to the polaronic shift of the
bottom of the band. The same polaronic effect introduces the
renormalization of the electron mass, $m_2$. There are two important
points worth mentioning here: (i) $\Sigma_2$ cannot introduce linear
in ${\bf p}$ corrections to the spectrum, and (ii) the sign of $m_2$
is unknown, we will assume that it is renormalized to a positive
value.

The parameters $Z_2$, $\mu_2$, and $m_2$ (recall that we are
discussing the case of $n_2=0$) are determined by the integration over
large momenta of the order of $p_{F1}$. That is why they can not be
calculated from the first principles and we treat them as input
parameters of the theory. If the interaction in the majority sector is
weak, then the calculation of $\mu_2$, and $m_2$ is possible. The
imaginary part of the remainder of the retarded self-energy can be
presented in the form (see Appendix~{\ref{A}})

\beq
\label{scalingsigma}
- \Im \tilde{\Sigma}_2 \left({\e}; p\right) =
\frac{\e\sqrt{2m_2|\e|}}{p_{F1}}
\left[f\left(\frac{p^2}{2m_2|\e|}\right) + 
{\cal O}\left(\frac{p}{p_{F1}}\right) \right],
\eeq 

\noindent
where $f(x)$ is a dimensionless function with properties

\beq
f(x) = \left\{ 
\begin{matrix}
{2}/{3} & x \to 0 \cr \cr
x^{1/2} & x \to \infty.
\end{matrix}
\right.
\eeq

\noindent
The above self-energy describes in particular the finite lifetime of
the minority electrons with respect to the emission of the
electron-hole pairs in the majority liquid. The rate of this decay is
proportional to $\e^{3/2}$. This is different from the usual
$\e^2\ln\e$ for the two dimensional Fermi liquid because there is no
Fermi surface for the minority electrons formed yet. However, the
quasiparticles are still well defined even in this case provided that
$\e \ll p_{F1}v_{F1}$. The form of Eq.~(\ref{scalingsigma}) follows
from simple dimensional analysis of corresponding diagrams which is
elaborated upon later in this section, however the dimensionless
function $f(x)$ can be obtained only by direct calculation, see
Appendix~\ref{A}.

As we already mentioned, the Green's function for the minority
electrons at $n_2=0$ is an analytic function of $\epsilon$ at the
upper semiplane ${\rm Im}\epsilon >0$. Therefore, the contribution of
two close poles is not dangerous and the singular part of the type of
Eq.~(\ref{twopoles}) does not arise.

Concluding this subsection, we emphasize that we have {\em assumed}
that the curvature of the spectrum at $k=0$ for the minority electrons
is positive. All the further scheme is based on this assumption, which
we cannot justify for arbitrary interaction strength. We will not
speculate on the alternative scenario in this paper.  For more
information on the minority Green's function and self-energy we refer
the reader to Appendix~\ref{A}.

\subsubsection{Vertex functions - general definitions}

To characterize interaction of the minority electrons
with each other as well as with the majority electrons
we will need $2n$-point vertex functions, which we denote by
$\Delta_{i_1\dots i_n}^{j_1\dots j_n}(\P_1^{out}, \dots \P_n^{out};
\P_1^{in}, \dots \P_n^{in})$, and define as 

\begin{widetext}
\beqa
&&-i
(2\pi)^3\delta\left(\sum_{k=1}^{n}\left(\P_k^{out}-\P_k^{in}\right)\right)
\Delta_{i_1\dots i_n}^{j_1\dots j_n}(\P_1^{out}, \dots \P_n^{out};
\P_1^{in}, \dots \P_n^{in}
)=\int d\R_1^{out} \dots d\R_n^{out}d\R_1^{in} \dots d\R_n^{in}
\label{Deltas}
\\
&&\times
\exp\left(-i\sum_{k=1}^{n}
\left(\P_k^{out}\R_k^{out}-\P_k^{in}\R_k^{in}\right)\right)
\langle T
\hat{\psi}_{j_1}(\R_1^{out})\dots\hat{\psi}_{j_n}(\R_n^{out})
\hat{\psi}_{j_1}^\dagger(\R_1^{in})\dots\hat{\psi}_{j_n}^\dagger(\R_n^{in})
\rangle^{amp}.
\nonumber
\eeqa
%\end{widetext} 
The external legs on each diagram are amputated.

Because of spin conservation, the number of incoming legs with
$j=1(2)$ equals to the number of outgoing legs with $j=1(2)$. Because
of the Fermi statistics the vertex function is antisymmetric with
respect to the permutations of outgoing legs
%\begin{widetext} 

\beqa
\Delta_{i_1\dots,i_k,\dots,i_l,\dots,i_n}^{j_1\dots j_n}
\left(\P_1^{out}, \dots, \P_{i_k}^{out},\dots,
\P_{i_l}^{out}
\dots,  \P_n^{out};
\P_1^{in}, \dots \P_n^{in}
\right)=
\nonumber\\
-
\Delta_{i_1\dots,i_l,\dots,i_k,\dots,i_n}^{j_1\dots j_n}
\left(\P_1^{out}, \dots, \P_{i_l}^{out},\dots,
\P_{i_k}^{out}
\dots,  \P_n^{out};
\P_1^{in}, \dots \P_n^{in}
\right);
\label{anti}
\eeqa

\noindent
\end{widetext} 
and have the same property for the ingoing ones.

Our aim is to identify the relation of the Fermi liquid constants with
the vertex functions in the theory. To do that we follow the standard
procedure of the Fermi liquid theory and explicitly separate those
contributions to vertex functions that contain possible
singularities. There are two sources of singularities: (i) overlaps of
poles of two Green's functions, see Eq.~(\ref{twopoles}), and (ii) the
Coulomb propagator. Exact vertex functions Eq.~(\ref{Deltas}) can be
built using $\Pi_1^{qp}$ and $V_0$ as the basic building blocks in
addition to the nonsingular part of $\Delta$:

\beq
\Gamma_{i_1\dots i_n}^{j_1\dots j_n}\equiv
\Delta_{i_1\dots i_n}^{j_1\dots j_n}\Big|_{\rm irreducible}.
\label{gammas}
\eeq

\noindent
Here ``irreducible'' means that $\Gamma$ comprises all the diagrams
that can not be cut by one Coulomb line or two majority Green's
functions. More precisely, in each diagram reducible in two
majority electrons we substitute only the smooth part of the
product of two corresponding Green's functions 

\beq \label{nopole} 
G_1 \left(\P +\frac{\Q}{2}
\right) G_1 \left(\P-\frac{\Q}{2}\right) \to \varphi(\P,\Q), 
\eeq 

\noindent
see Eq.~(\ref{twopoles}). Irreducible vertices (\ref{gammas}) obey the
antisymmetry relation (\ref{anti}).

The total vertex function can be quite easily found from the
irreducible ones. The corresponding relation for the 4-point vertex
function is presented on Fig.~\ref{vertex}. In this scheme the Fermi liquid 
parameters are going to be determined by the vertex $W$.

The 3-point irreducible (in the same sense as $\Gamma$) vertex
function $B_{i}(\P,\Q)$ satisfies the Ward identity

\beq \label{WardB}
B_{i}(\P,0)= 1- \frac{\partial \Sigma_i(\e,\p)}{\partial \e},
\eeq

\noindent
where $\P=(\e,\p)$. Notice that due to the irreducibility definition,
see Eq.~(\ref{nopole}), the question of the order of limits is
resolved automatically.

{ 
\begin{figure}[ht]  
%\vspace{0.1 cm} 
\epsfxsize=8.2 cm  
\centerline{\epsfbox{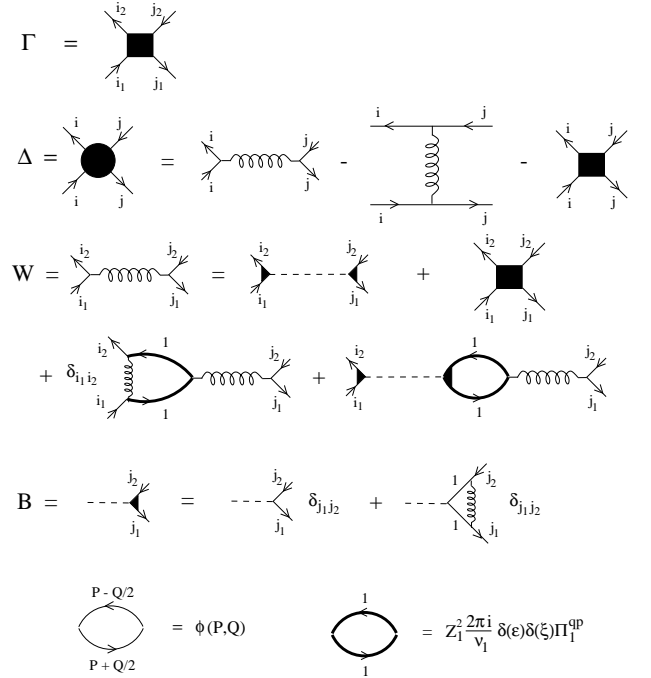}} 
\vspace{0.2cm} 
\caption{Relation of irreducible ($\Gamma$, $B$) and reducible
($\Delta$) vertices.}
\label{vertex} 
\end{figure} 
}

At $n_2=0$ any closed loop for the minority electrons vanishes, so
majority electrons obey the standard Fermi liquid description, which
does not depend on the values of $\Gamma_{12}^{12}$ and
$\Gamma_{22}^{22}$.  Calculation of vertices involving minority
electrons, however, requires knowledge of the irreducible vertex
functions $\Gamma_{12}^{12}$ and $\Gamma_{22}^{22}$; we will need
their values at external momenta much smaller than $p_{F1}$. We intend
to prove the smallness of $\Gamma_{22}^{22}$ and determine the
dependence of $\Gamma_{12}^{12}$ on external momenta in this region.
The proof is based on the dimensional analysis of each order of the
perturbation theory for the minority vertices.

In building further perturbation theory for the finite density of
minority electrons, we will use the screened interaction (\ref{Vqw})
as the basic interaction propagator, because it already contains all
the singularities (\ref{twopoles}) of the theory. In particular,
$V(\w=0,q)$ is always finite and short range, unlike the bare
interaction.  We will see later that the relevant contributions come
from $\w \ll v_{F1}q$; in this region, we can easily solve
Eq.~(\ref{PiFermi}), and obtain from Eq.~(\ref{Vqw})

\beqa 
V(q,\w) &\approx& \frac{1}{\nu_1} \left[ 1+F_{11}^{(0)} -
  \frac{\left(1+F_{11}^{(0)}\right)^2}{\nu_1 V_0(q)}
  -\frac{i|\w|}{qv_{F1}} + \dots
\right] 
\nonumber\\
&&
\nonumber\\
&=& \frac{1}{\nu_1} \left[ 1 +F_{11}^{(0)} - \frac{q}{q_s}
  -\frac{i|\w|}{qv_{F1}} + \dots \right].
\label{vaprox}
\eeqa

\noindent
The wavevector
 
\beq
q_s \equiv  
p_{F1}\left(\frac{e^2}{v_{F1}}\right)
\frac{1}{\left(1+F_{11}^{(0)}\right)^2}
\eeq 

\noindent
characterizes the screening of the Coulomb potential by majority
electrons. For reasonable interaction strength $q_s$ is not that
different from $p_{F1}$. That is why we will not write ratio
$q_s/p_{F1}$ in the subsequent estimates unless it is necessary for a
quantitative analysis.

\subsubsection{Vertex functions $\Gamma$ for minority electrons}

Let us start with the vertex function involving ingoing and outgoing
legs for minority electrons. We intend to show that the form of the
potential (\ref{vaprox}) and the antisymmetry relation (\ref{anti}),
guarantees that for the small external energies and momenta, $p_{i}
\ll p_{F1},\ \e_{i} \ll v_{F1}p_{F1}$, the $2n$ point function has the
following structure:

\beqa
&&\Gamma_{2,\dots,2}^{2,\dots,2}(\P_1^{out}, \dots, \P_n^{in}
)\nonumber\\
&&\quad=\frac{\gamma^{(n)}\left(\left\{\frac{\p_i}{Q}\right\}
, \left\{\frac{2m_2\e_i}{Q^2}\right\}
\right)}{\nu_2 \left(p_{F1}\right)^nQ^{n-4}}
\left[1
+ {\cal O}\left(\frac{Q}{p_{F1}}\right)
\right];
\nonumber\\
&&Q\equiv  \left[\sum_{i=1}^n
\left([\p_i^{in}]^2+[\p_i^{out}]^2\right)\right]^{1/2},
\label{scaling}
\eeqa where $\gamma^{(n)}$ is a finite dimensionless function, obeying
the antisymmetricity relation following from Eq.~(\ref{anti}).

Relation (\ref{scaling}) can be shown as following. Consider any order
of the perturbation theory (lowest non-vanishing diagrams for $4-$ and
$6-$ point vertices are shown in Fig.~\ref{G22lowest}). We notice that
there are two scales in the problem.  The first, ``ultraviolet'' scale
is determined by the majority electrons, i.e. the wavevectors of the
order of $p_{F1}$ and the energies of the order of $v_{F1}p_{F1}$. The
second, ``infrared'' scale is determined by the momenta and energies
of the external legs, i.e. the scale of the integration over the
momentum and energy is qiven by $Q$ and $Q^2/(2m_2)$ respectively.
Statement (\ref{scaling}) is the obvious consequence of the
perturbation theory for infrared diagrams. Indeed, the lowest order
nonvanishing diagram for the $2n$-point function contains $n$
interaction lines and $n$ electron minority Green functions, see
Fig.~\ref{G22lowest}.  Calculating a diagram in this regime one can
use approximation (\ref{vaprox}) for the interaction potential. The
constant part of the potential (corresponding to contact interation)
cancels immediately from the whole theory (i.e. from any vertex
function $\Delta$) when being substituted in any interaction line, see
e.g. Fig.~\ref{F22second}., due to the antisymmetry relation
(\ref{anti}), which is a simple manifestation of the fact that
spinless fermions are not sensitive to the contact interaction.  Only
does the remaining part of the interaction

\[
\delta V(q,\w)=-\frac{1}{\nu_1}
\left[ \frac{q}{q_s} + \frac{i|\w|}{qv_{F1}}
\right] 
\]
contribute to the final answer. Then any infrared integral like

\[
{\cal I}_n=\int d^2 p d \e [\delta V(p,\e)]^n [G_2(p,\e)]^n
\] 
[we omit external momenta here] can be made dimensionless by
expressing all momenta in units of $Q$, and all energies in units of
$Q^2/ 2 m_2$. This way, we obtain

\[
{\cal I}_n
=Q^2 \frac{Q^2}{2m_2} \left(\frac{Q^2}{2m_2}\right)^{-n}
\left(\frac{Q}{m_2 p_{F1}}\right)^n  \times
\begin{pmatrix} {\rm dimensionless}\cr  {\rm function} \end{pmatrix}, 
\]
which clearly has the form of Eq.~(\ref{scaling}).
Inclusion of additional $m$ interaction lines bearing small momenta 
into the tree level diagrams in Fig.~\ref{G22lowest}, provides additional
smallness $(Q/p_{F1})^m$.

{ 
\begin{figure}[t]  
%\vspace{0.1 cm} 
\epsfxsize=6 cm  
\centerline{\epsfbox{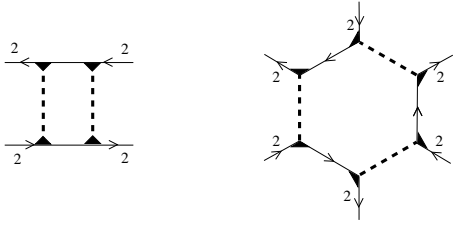}} 
\vspace{0.2cm} 
\caption{Lowest order contributions to 4- and 6-point minority vertex
functions $\Gamma$.  Dashed lines denote the non-constant part of the
screened Coulomb interaction $\delta V$, internal lines are minority
propagators $G_2$.  Black triangles denote the vertex $B$, see
Fig.~\ref{vertex}.}
\label{G22lowest} 
\end{figure} 
} 

This procedure of finding the scaling form of the minority vertex
function $\Gamma$ relies on assumption that the integrals are
determined by the small momentum region.  To justify this assumption,
let us show that the contribution from the ultraviolet is always
small.  Indeed, let us separate the contribution into $\Gamma$ from
Eq.~(\ref{scaling}) where all the integrals are determined by the
ultraviolet parts. Then we can introduce the momentum scale $k^\ast =
A p_{F1}$, where $A$ is a numerical coefficient smaller than $1$, and
restrict integration over momenta by $k>k^\ast$ and over the energy by
$|\epsilon| > v_{F1}k^\ast$, and call this contribution
$\tilde\Gamma$. Because the integrals are restricted to the high
momentum region, $\tilde\Gamma$ is an analytic function of its
external momenta and energies and can be expanded in Taylor
series. Because of the antisymmetricity constraint (\ref{anti}), the
first nonvanishing term has the form (put all external energies to $0$
for the sake of simplicity):

\beqa
&&\tilde{\Gamma}_{2,\dots,2}^{2,\dots,2}(\P_1^{out}, \dots, \P_n^{in}
) \simeq \frac{1}{\left[p_{F1}\right]^{2n-4}\nu_1 }
\label{ultraviolet}
\\
&& \quad \times
\frac{
\prod_{i<j}\prod_{k<l}\left(\p^{(in)}_i - \p^{(in)}_j\right)
\cdot \left(\p^{(out)}_k - \p^{(out)}_l\right)
}{\left[k^{\ast}\right]^{n (n-1)}}.
\nonumber  
\eeqa

\noindent
Because, by construction, $k^\ast$ is different from $p_{F1}$ only by
a numerical factor, this estimate is smaller than the result of
Eq.~(\ref{scaling}) by a factor of $\left(Q/p_{F1}\right)^{n^2-4}$,
for $n>2$.

The case of the $4$-point vertex is special -- ultraviolet and
infrared estimates have the same powers of $p_{F1}$ in the
denominator.  It means that this vertex function is uniformly
contributed by all energy scales. Therefore, function $\gamma^{(2)}$,
may contain logarithmic dependence of the high energy scale $p_{F1}$.
Indeed, the direct calculation of this particular vertex shown in
Appendix~\ref{B} gives

\beqa
\Gamma_{2,2}^{2,2}&&\left(\P_1^{out}, \P_2^{out};\P_1^{in}, \P_2^{in}\right)
= \label{gamma22log} \\
&& \frac{1}{\nu_2} \frac{Q^2}{[p_{F1}]^2}
\left[\gamma_1 \ln \frac{Q}{p_{F1}} + \gamma_2
+ {\cal O}\left(\frac{Q}{p_{F1}}\right)\right],
\nonumber
\eeqa

\noindent
where the dimensionless functions $\gamma_i$ describe the dependence
on the external momenta and energies: $\gamma_i =
\gamma_i\left(\left\{\frac{\p_j}{Q}\right\} ,
\left\{\frac{2m_2\e_j}{Q^2}\right\}\right)$.  From the dimensional
analysis above, it follows that all the leading graphs should have one
infrared loop similar to the tree level diagrams.  All other loops
must be ultraviolet -- their role is the ``dressing'' of the vertices
of the tree-level diagrams [black triangles in Fig.~\ref{G22lowest}].
This dressing changes the numerical coefficient in the final
expressions but does not affect the analytic structure of
Eq.~(\ref{scaling}).

\subsubsection{Vertex functions $\Gamma$ involving minority and 
majority electrons}

The next object to consider is the vertex function involving both
minority and majority spins. We start from the simplest vertex
$\Gamma_{12}^{12}$ with two minority and two majority legs.  This
object characterizes the correction to the simple Coulomb interaction
between minority and majority electrons, the lowest order diagrams
contributing to this vertex are shown on Fig.~\ref{G12lowest}.  We
will be interested in the behavior of this vertex when the energy and
momentum transfers are small in comparison with the Fermi energy and
Fermi momentum of majority electrons, but are arbitrary in comparison
with energy and momentum of minority electrons. Furthermore, we are
interested in the situation, where the majority legs are nearly
on-shell. We intend to show, that this vertex has the form

\beqa
&&{\nu_1}\Gamma_{12}^{12}(\P_1^{out},\P_2^{out};\P_1^{in},\P_2^{in})
= \gamma_{12}^{(2)} + \frac{\p_1 \cdot\p_2}
{\left[p_{F1}\right]^2} \eta_{12}^{(2)} 
\label{Gamma12} 
\\
&& \quad \quad + {\cal O}\left(\frac{\e}{\e_{F1}}\right)+ 
{\cal O}\left(\frac{\xi_1}{\e_{F1}}\right)+ 
{\cal O}\left(\left[\frac{\p_2}{p_{F1}}\right]^2\right).
\nonumber\\
&&\P_{1,2}^{in}=\left(\e_{1,2}\pm\frac{\w}{2},\ 
  \p_{1,2}\pm\frac{\k}{2}\right); \nonumber \\ &&
\P_{1,2}^{out}=\left(\e_{1,2}\mp\frac{\w}{2},\
  \p_{1,2}\mp\frac{\k}{2}\right), \ \ 
\xi_1=v_{F1}\left(|\p_1|-p_{F1}\right)
\nonumber
\eeqa

\noindent
where $\gamma_{12}^{(2)}$ and $\eta_{12}^{(2)}$ are finite numerical
coefficients.

To understand the relation (\ref{Gamma12}), we notice that it is
equivalent to the statement that $\Gamma_{12}^{12}$ can be expanded in
a Taylor series as a function of momenta and energy of the minority
electrons. The form of the term linear in $\p_2$ is guarded by the
rotational and time reversal symmetries. What we need to prove is that
the Taylor expansion indeed exists. This would be true if the dominant
contribution to the vertex came from the ``ultraviolet'' region [in
the same sense asused for derivation of Eq.~(\ref{ultraviolet})].

The only suspicious region where the non-analytic dependence of $\P_2$
can arise is the infrared integration in the diagrams which are
reducible in one minority and one majority line, see
Fig.~\ref{G12lowest}. (a) and (b). Indeed, only in this case is the
appearance of the overlaping poles possible, which may lead to the
nonalyticity. Let us, however, examine the expressions for both those
diagrams in more detail. Let us write their analytic expression:

\beqa
&& {\rm Fig}.~\ref{G12lowest}. (a)+{\rm Fig}.~\ref{G12lowest}. 
(b)\simeq \int d\Omega d^2q\times
\nonumber\\
&& \times
V\left(\q_+,\W_+\right)
V\left(\q_-,\W_-\right)G_2\left(\e_2-\W,\p_2 -\q\right))
\nonumber\\
&&\times
\left[
G_1\left(\e_1+\W,\p_1 +\q\right)
+G_1\left(\e_1-\W,\p_1 -\q\right)
\right]
\nonumber\\
&&\q_\pm =\q\pm\frac{\k}{2}, \quad \W_\pm =\W\pm\frac{\w}{2}
\label{F12estim}
\eeqa

\noindent
where the screened potential is given by Eq.~(\ref{Vqw}).  The
dangerous contribution may come only from the pole part of the
majority Green'S function (\ref{gf1}). For $|p|=p_{F1}$ we have

\beq \label{ehsymm}
G_1\left(\e,\p +\q\right)= -
G_1\left(-\e,\p_1 - \q\right),
\eeq

\noindent
which is nothing but the electron-hole symmetry for linearized
spectrum. Therefore, the two terms in brackets in Eq.~(\ref{F12estim})
cancel each other at $\e_1,\xi_1 = 0$, and therefore no infrared
contribution is possible.  All the corrections associated with the
electron- hole asymmetry of majority electrons have at least one extra
power of the Fermi energy $\e_{F1}$ in the denominator.

It is clear that the same argument about the electron-hole symmetry
remains valid even if the interaction lines $V$ are replaced by the
dressed vertices (\ref{Gamma12}), and therefore, the form
(\ref{Gamma12}) persists in all of the orders of the perturbation theory.

{ 
\begin{figure}[t]  
%\vspace{0.1 cm} 
\epsfxsize=6 cm  
\centerline{\epsfbox{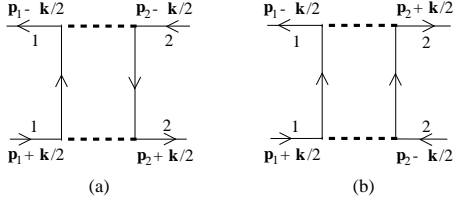}} 
\vspace{0.2cm} 
\caption{Lowest order diagrams for $\Gamma_{12}$.} 
\label{G12lowest} 
\end{figure} 
}

So far, we established that the infrared part of the diagrams
involving mutual scattering of one minority and one majority electrons
does not contribute because of the electron-hole symmetry of the
majority system. It is clear that the above reasoning can be applied
to higher order vertices as well. For further analysis, we will need
only the $6$-point vertex involving one majority and two minority
electrons.  Repeating all of the above consideration and taking into
account the antysimmetricity (\ref{anti}) with respect to the
permutations of the minority electrons, we find with logarithmic
accuracy (the majority electrons are assumed to be on shell):

\begin{widetext} 
\beq
{\nu_1}\Gamma_{122}^{122}\left(\P_1^{out},\P_2^{out}, \P_3^{out};
\P_1^{in},\P_2^{in}, \P_3^{in}\right)
=
\frac{\left(\p_2^{in}- \p_3^{in}\right)\cdot
\left(\p_2^{out}- \p_3^{out}\right) }{\left[p_{F1}\right]^4}
\left[\gamma_{122}^{(3)}\ln\left(\frac{p_{F1}}
{{\rm max}(p_2^{in}, p_3^{in};\ p_2^{out}, p_3^{out} )}\right)
+ {\cal O} (1)\right],
\label{Gamma122} 
\eeq
\end{widetext} 

\noindent
where $\gamma_{122}^{(3)}$ is the coefficient of the order of unity;
we will not need its value in the subsequent consideration.
This formula may be understood as the dependence of the prefactor in
the two-particle vertex function 
(\ref{gamma22log}) on the density of the majority electrons.

\subsubsection{Vertex functions W}

The established dependence of the vertex functions $\Gamma_{12}^{12};
\Gamma_{22}^{22}$ enables us to find the dependence of functions $W$
from Fig.~\ref{vertex}. explicitly.  We will see later that the value
of $W$ on shell is directly related to the Fermi liquid constants.  We
solve the diagramatic equation in Fig.~\ref{vertex}. at $q \ll
p_{F1}$, and $\w \ll v_{F1}p_{F1}$.  Under this condition, we may
replace $B_i(\P,\Q) \to B_i(\P,0)$ and use the Ward identity
(\ref{WardB}).  Moreover, on majority shell $B_1(\P,0) = 1/Z_1$ and
for $\e \ll v_{F1}p_{F1}$ and $p \ll p_{F1}$, $B_2(\P,0) = 1/Z_2$.
This yields at $\w \ll v_{F1}q$

\begin{widetext} 

\beq
W_{22}^{22}(\P_1,\P_2; \w, \q) 
= \frac{1}{Z_2^2\nu_1}
  \left[ 1+F_{11}^{(0)} 
  -\frac{i|\w|}{qv_{F1}}\right] -\frac{2 Z_1\gamma_{12}}{Z_2\nu_1}
 - \left[ \frac{\left( 1+F_{11}^{(0)} \right)^2 }{Z_2^2} 
- \frac{2 Z_1\gamma_{12}\left( 1+F_{11}^{(0)} \right) }{Z_2} + 
\left(Z_1\gamma_{12}\right)^2 \right] \frac{1}{\nu_1^2 V_0(q)}. 
\label{Ws}
\eeq
\end{widetext} 

\noindent
The leading correction to Eq.~(\ref{Ws}) comes from
Eq.~(\ref{gamma22log}) and it has the estimate
$(\p_1/p_{F1})^2\ln(\p_{1}^2/p_{F1}^2)$. The momentum dependence of
the vertices $B$ give a correction of the order of
$(\p_1/p_{F1})^2$. We neglect the corrections of this kind.

The other function we need in further calculations is
$W_{12}^{12}(\P_1,\P_2; \w, \q)$.  The Coulomb interaction line does
not flip the spin of the electron and therefore this object is
determined solely by the irreducible vertex (\ref{Gamma12}). We find

\beq
W_{12}^{12}(\P_1,\P_2;\w, \q)
 = -
\frac{\gamma_{12}^{(2)}}{\nu_1} - \frac{(\p_1+\p_2)^2-\q^2}
{\nu_1\left[p_{F1}\right]^2} \eta_{12}^{(2)}. 
\label{W12}
\eeq

\noindent
The minus sign here is associated with the change of the direction at
which the irreducible vertex enters the diagram.

In the following subsection we will use Eq.~(\ref{Ws}) to find the
value of the Fermi liquid parameters.

\subsection{System near full polarization}
\label{3c}

In this subsection we will take the following route. First, we
calculate the changes in the Green function due to the finite density
minority electrons but considering their spectrum unchanged at
$n_2=0$. Second, we compute change in the vertex functions due to the
finite density. Finally, we use the vertex functions to recalculate
the spectrum of the minority and majority electrons, thus determining
the Fermi liquid function. The small parameter justifying the
procedure is $n_2/N \ll 1$.

\subsubsection{Green functions}

For nonzero minority electron density $n_2 > 0, \ n_2 \ll n_1$ we
begin our discussion by considering the minority Green function
$\tilde{G}_2$ (hereinafter Green functions and n-point functions with
tilde are understood at $n_2 >0$, while the absence of the tilde
implies $n_2=0$). Finite density of majority electrons leads to the
appearance of the positive chemical potential $\tilde{\mu}_2$ and the
shift of the pole in Eq.~(\ref{g2}) to the upper semiplane, at $\rm Re
\ \e < 0$.  This change is described as

\begin{subequations}
\label{G2new}
\beq
\tilde{G}_2(\e, \p) = G_2(\e + \tilde{\mu}_2, \p) + \delta G_2(\e,
\p) + \delta G_2^{sm}(\e,\p).
\label{G2change}
\eeq

\noindent
The second term in Eq.~(\ref{G2change}) originates from the
quasiparticle pole of the Green function

\beq
\delta G_2 = i 2\pi Z_2 \delta 
\left(\e + \tilde{\mu}_2 - \frac{p^2}{2 {m}_2}\right) 
\theta \left( p_{F2} - p \right), 
\label{G2pole}
\eeq

\noindent
where $\theta(x)$ is the Heaviside step-function. Here we neglected
the correction to the parabolic spectrum and will restore this
dependence later on. Within this approximation, $\tilde{\mu}_2 =
{p^2_{F2}}/{2 {m}_2}$.

The last term in Eq.~(\ref{G2change}) gives zero while integrated over
$\e$ only in the vicinity of the minority Fermi surface. However, this
term is not an analytic function of $\e$ in the upper semiplane and
therefore it gives finite contribution to the density.  In the leading
order in $n_2$ it can be found from Fig.~\ref{dS2} and equals to

\beqa
&&\delta G_2^{sm}(\e,\p)= \nonumber\\
&&\quad
i\left[G_2(\e,\p)\right]^2  \int  
\Gamma_{22}^{22} \left( \P, \P_1 ; \P, \P_1 \right) \delta G_2(\P_1) 
\frac{d^3 \P_1}{(2\pi)^3}\nonumber\\
\nonumber\\
&&\quad
=-n_2 Z_2 \Gamma_{22}^{22} 
\left( \P, \0 ; \P, \0 \right)\left[G_2(\e, \p)\right]^2.
\label{G2smooth}
\eeqa

\noindent
In this expression we implied that only $p \gg p_{F2}$ will contribute
to the observable quantities and that is why we put two argument of
the vertex function to zero. For the same reason, the chemical
potential $\tilde{\mu}_2$ can be neglected in the argument of the
Green function.

\end{subequations}

The presence of the smooth term (\ref{G2smooth}) is crucial for the
gauge invariance of the theory. In particular, it provides the
cancellation of gauge-noninvariant factor $Z_2$ from the observable
quantities.  As an example, we calculate the electron density
$n_2$. Using Eqs.~(\ref{G2new}) and the fact that $G_2(\e)$ is
analytic for ${\rm Im} \e > 0$, we find

\beqa
n_2 &=& -i \int \frac{d^3 \P}{(2\pi)^3} e^{i0\e} G_2(\P)
= Z_2\int\frac{d^2p}{(2\pi)^2}\theta \left( p_{F2} - p \right)
\nonumber\\
&&+i n_2 Z_2 \int \frac{d^3 \P}{(2\pi)^3} 
\Gamma_{22}^{22} \left( \P, \0 ; \P, \0 \right)G_2(\e, \p)^2 
\nonumber 
\\
&&  = \frac{ Z_2 p_{F2}^2}{4\pi} - n_2 \left(Z_2-1\right),
\label{noZ2}
\eeqa 

\noindent
where in the last transformation we used the Ward identity
(\ref{WardB}), and

\[
\frac{\partial \Sigma_2}{\partial \e}\Big|_{\e \to 0}=1-\frac{1}{Z_2}.
\]

\noindent
It is seen from from Eq.~(\ref{noZ2}) that the quasiparticle weight
$Z_2$ is cancelled and the Landau theorem $n_2=p_{F2}^2/4\pi$ holds.

{ 
\begin{figure}[t]  
%\vspace{0.1 cm} 
\epsfxsize=2 cm  
\centerline{\epsfbox{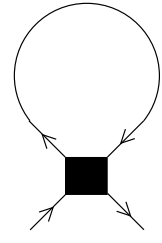}} 
\vspace{0.2cm} 
\caption{Contribution to $\delta \Sigma_2$.
The internal line denotes $\delta G_2$.}
\label{dS2} 
\end{figure} 
}

\subsubsection{Vertex functions: Fermi liquid parameters and 
their corrections.}

To calculate the Fermi liquid constant $F_{ij}(\widehat{\nn_1\nn_2})$
one may start from the standard expression, see Fig.~\ref{Fij}.

\beqa
&&F_{ij}(\widehat{\nn_1\nn_2}) = -Z_iZ_j\nu_1W_{ij}^{ij}
\left(\P; \P; \q; \w =0 
  \right);
\label{FLdef}
\\
&&\P=
\left(\frac{p_{Fi}\nn_1 + p_{Fj}\nn_2}{2},\e=0\right),
\quad
\q=p_{Fi}\nn_1 - {p_{Fj}\nn_2},
\nonumber
\eeqa

\noindent
where the factor $\nu_1$ is introduced in order to make the constants
dimensionless.  Notice, that with the vertex functions defined as in
the previous section, the usual problem on non-commuting limits $\w
\to 0; \q \to 0$ does not arise at all.

{ 
\begin{figure}[t]  
%\vspace{0.1 cm} 
\epsfxsize=6 cm  
\centerline{\epsfbox{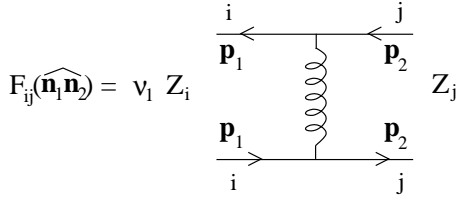}} 
\vspace{0.2cm} 
\caption{Definition of Fermi-liquid constants.}
\label{Fij} 
\end{figure} 
} 

For the majority electrons Eq.~(\ref{FLdef}) is just a formal
definition which does not bring anything new. Actually, this
definition was already used in the derivation of Eqs.~(\ref{Vqw}) and
(\ref{v0}). For the minority electrons, Eq.~(\ref{FLdef}) actually
allows one to calculate the Fermi liquid parameters.  Substituting
Eq.~(\ref{W12}) into Eq.~(\ref{FLdef}) we find

\beq
F_{12}
\left(\widehat{\nn_1 \nn_2}\right) = 
Z_1Z_2\left(\gamma_{12}^{(2)}+ \frac{p_{F2}}{p_{F1}}
\eta_{12}^{(2)}\nn_1\cdot \nn_2\right).
\label{F12}
\eeq

In order to find $F_{22}$ we substitute Eq.~(\ref{Ws}) into
Eq.~(\ref{FLdef}). Using zero angular harmonics of Eq.~(\ref{F12}) to
eliminate constant $\gamma_{12}^{(2)}$, we obtain Eq.~(\ref{F22ang}).

The derivation presented here is still too cavalier. The expression
(\ref{F22ang}), for instance, contains the dependence on the density
of the majority electrons. However, this dependence came solely from
the momentum dependence of the vertex functions $W_{22}^{22}$ which
was calculated at $n_2=0$. Finite density $n_2$ leads to the
contributions from the closed loops of the minority electrons and thus
to modification of $W_{22}^{22}$. To prove the legitimacy of retaining
the second term in Eq.~(\ref{F22ang}), one has to prove that the
modification of $W_{22}^{22}$ is higher order in $n_2/N$. We turn to
the corresponding proof now.
 
{ 
\begin{figure}[t]  
%\vspace{0.1 cm} 
\epsfxsize=7 cm  
\centerline{\epsfbox{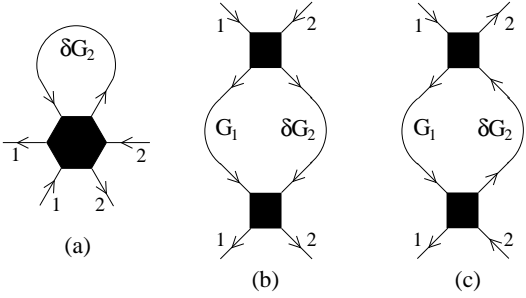}} 
\vspace{0.2cm} 
\caption{Leading contributions to $F_{12}$. The filled hexagon denotes 
the irreducible vertex $\Gamma_{122}^{122}$.}
\label{G12pert} 
\end{figure} 
}

\subsubsection{Corrections to minority-majority Fermi liquid 
parameter $F_{12}$ }

The correction to the Fermi-liquid constant $F_{12}$ comes only from
the irreducible vertex. This correction is shown on
Fig.~\ref{G12pert}. (a)-(c).  Contribution due to the irreducible
6-point vertex is given by

\beqa
&& {\rm Fig.~\ref{G12pert}.}{\rm (a)} = \nonumber \\
&& \quad \int \frac{d^3\P}{(2\pi)^3} 
\delta G_2\left(\P \right)
 \Gamma_{122}^{122}\left(\P_1^{out},\P_2^{out}, \P;
\P_1^{in},\P_2^{in}, \P\right). 
\nonumber
\eeqa

\noindent
Using Eqs.~(\ref{Gamma122}) and (\ref{G2pole}), we arrive to the estimate

\[
{\rm Fig.~\ref{G12pert}.(a)} \propto \left(n_2\right)^2,
\]

\noindent
and this contribution is negligible.  Contributions of
Fig.~\ref{G12pert}. (b) and (c) are not small separately but their sum
is.  Presense of $\delta$-function in Eq.~(\ref{G2pole}) guarantees
the momentum and energy transfer to be small, therefore, the expansion
(\ref{Gamma12}) for the vertex functions is legitimate. We thus find

\beqa
&& {\rm Fig.~\ref{G12pert}}. {\rm (b)} +  {\rm Fig.~\ref{G12pert}}.{\rm (c)}
=\left(\frac{\gamma_{12}}{\nu_1}\right)^2
\int \frac{d^3\P}{(2\pi)^3}\delta G_2\left(\P \right)\nonumber \\
&& \times \left[G_1 \left( \e, \p +\q \right) 
+ G_1 \left( -\e, \p-\q \right) \right].  
\eeqa

\noindent
Contribution from the pole parts of the majority Green functions is
cancelled due to the electron-hole symmetry, compare with
Eq.~(\ref{F12estim}), and we obtain the finite contribution
proportional to at least the first power of $n_2$.

\subsubsection{Corrections to minoroty Fermi liquid parameter $F_{22}$}

Similarly, corrections to the Fermi liquid constant $F_{22}$ are
determined by the six-point irreducible vertex, see
Fig.~\ref{G22pert}. (a) and by the sum of six graphs determined by
$4$-point vertices $W_{22}^{22}$ from Eq.~(\ref{Ws}), see
Fig.~\ref{G22pert}. (b)-(g).  Contribution of the first graph is
estimated with the help of Eq.~(\ref{scaling}) for $n=3$ and
Eq.~(\ref{G2pole}):

\begin{eqnarray*}
&&{\rm Fig.~\ref{G22pert}.}{\rm (a)} = \\
&&\quad \int \frac{d^3\P}{(2\pi)^3} 
\delta G_2\left(\P \right)
 \Gamma_{222}^{222}\left(\P_1^{out},\P_2^{out}, \P;
\P_1^{in},\P_2^{in}, \P\right)\\
&&\quad \simeq  \frac{\left(\p_1^{in}- \p_2^{in}\right)\cdot
\left(\p_1^{out}- \p_2^{out}\right) }{\nu_1[p_{F1}]^2}, 
\end{eqnarray*}

\noindent
where the external momenta are assumed to be on shell.  This result
can be understood as the correction to the argument of the logarithm
in Eq.~(\ref{gamma22log}).  Because we already neglected the term
(\ref{gamma22log}) in Eq.~(\ref{Ws}) as producing $(n_2/N)\ln(n_2/N)$
correction, keeping the correction of Fig.~\ref{G22pert}.(a) would be
also beyond the accuracy of the calculation and we can neglect it.

Each term in the reducible graphs Fig. (b) - (e) is not small but
their sum is.  One finds

\begin{widetext} 
\beqa
&&{\rm Fig.~\ref{G22pert}.} {\rm (b)} + \dots + 
{\rm Fig.~\ref{G22pert}.} {\rm (e)}=
\int \frac{d^3 \Q}{(2\pi )^3}
\left[\delta G_2(\P_2+\Q)  G_2(\P_1+\Q) + G_2(\P_2+\Q)  
\delta G_2(\P_1+\Q)\right] \nonumber\\ 
&&\quad \times
\left[W_{22}\left(\P_1+\frac{\Q}{2}, \P_2+\frac{\Q}{2}; \Q \right)
- W_{22}\left(\Q+ \frac{\P_1+\P_2}{2}, 
\frac{\P_1+\P_2}{2}; \P_2-\P_1 \right)
\right]^2. \label{be}  
\eeqa
\end{widetext} 

\noindent
The Green function $\delta G_2$ from Eq.~(\ref{G2pole}) restrict the
integration in the infrared region, where one can use Eq.~(\ref{Ws})
for function $W_{22}$, and Eqs.~(\ref{g2}) and (\ref{sigma2}) for the
Green function $G_2$. The constant part of $W_{22}$ immediately
cancels, and we repeat the dimensional analysis which lead us to
estimate (\ref{scaling}). This immediately yields the result
proportional to $n_2$, which can be neglected.

The last two diagrams are analyzed in the same manner:

\begin{widetext} 
\beqa
&&{\rm Fig.~\ref{G22pert}.} {\rm (f)} + {\rm Fig.~\ref{G22pert}.} {\rm (g)}=
\int \frac{d^3 \Q}{(2\pi )^3} {\cal I}(\P_1; \P_2; \Q) 
\nonumber
\\
&&{\cal I}(\P_1; \P_2; \Q)=
\left[\delta G_2(\P_2+\Q)  G_2(\P_1-\Q) + G_2(\P_2+\Q)  
\delta G_2(\P_1-\Q)\right] 
W_{22}\left(\P_1+\frac{\Q}{2}, \P_2+\frac{\Q}{2}; \Q \right)\nonumber\\ 
&&\quad \times
\left[W_{22}\left(\P_1+\frac{\Q}{2}, \P_2+\frac{\Q}{2}; \Q \right)
  - W_{22}\left( \frac{\P_1+\P_2}{2} - \Q, 
\frac{\P_1+\P_2}{2} + \Q; \Q+ \P_1-\P_2 \right)
\right]. \nonumber   
\eeqa

\noindent
Shifting the integration variable, we replace ${\cal I}(\P_1; \P_2;
\Q) \to [{\cal I}(\P_1; \P_2; \Q) + {\cal I}(\P_1; \P_2;
\Q+\P_2-\P_1)]/2$ in the integrand and find

\beqa
&&{\rm Fig.~\ref{G22pert}.} {\rm (f)} + {\rm Fig.~\ref{G22pert}.} 
{\rm (g)}=\frac{1}{2}
\int \frac{d^3 \Q}{(2\pi )^3} 
\left[\delta G_2(\P_2+\Q)  G_2(\P_1-\Q) + G_2(\P_2+\Q)  
\delta G_2(\P_1-\Q)\right] 
\nonumber\\
&& \times
\left[W_{22}\left(\P_1+\frac{\Q}{2}, \P_2+\frac{\Q}{2}; \Q \right)
  - W_{22}\left( \frac{\P_1+\P_2}{2} - \Q, \frac{\P_1+\P_2}{2} 
+ \Q; \Q+ \P_1-\P_2 \right)
\right]^2. 
\nonumber   
\eeqa
\end{widetext} 

\noindent
Similarly to Eq.~(\ref{be}), it yields the result proportional to the
first power of $n_2$.

{ 
\begin{figure}[t]  
%\vspace{0.1 cm} 
\epsfxsize=8 cm  
\centerline{\epsfbox{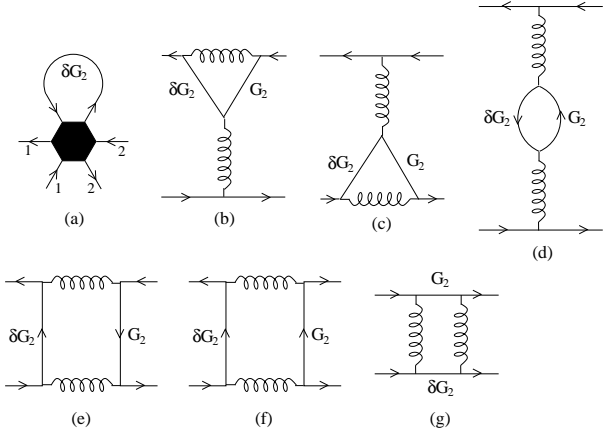}} 
\vspace{0.2cm} 
\caption{Leading contributions to $F_{22}$. The filled hexagon denotes
the irreducible vertex $\Gamma_{222}^{222}$. For each of the diagrams
(b)-(g) there exists a counterpart obtained by swapping $\delta G_2$
and $G_2$.}
\label{G22pert} 
\end{figure} 
}

\subsubsection{Final remarks}

Let us now discuss the effect of the finite density of minority
electrons on the majority electrons. Firstly, one observes that the
minority polarization operator is not small: $\Pi_2=\nu_2$ at momentum
transfer $q < 2 p_{F2}$ and decays as $\Pi_2 \simeq \nu_2
(p_{F2}/q)^2$ at larger momenta. Naive calculation of the correction
to the majority Fermi liquid parameter $F_{11}(\theta)$ gives large
contribution to the forward scattering for angles $\theta <
p_{F2}/p_{F1}$.  This contribution, however, is cancelled out from any
closed loop for the majority electrons, see Fig.~\ref{loop}.  The
easiest way to see it, is by noticing that any closed loop with
arbitrary number of scalar vertices vanishes due to the gauge
invariance if one of the momenta equals to zero.  Therefore, one may
gauge away those contributions from the very beginning, and leave only
processes with momentum transfer of the order of $p_{F1}$. In this
region, correction to the majority loops due to the presence of a
nonvanishing density of minority electrons is already small as
$[p_{F2}/p_{F1}]^2$, which ones again results in the correction
proportional to the first power of $n_2$.

{ 
\begin{figure}[t]  
%\vspace{0.1 cm} 
\epsfxsize=7 cm  
\centerline{\epsfbox{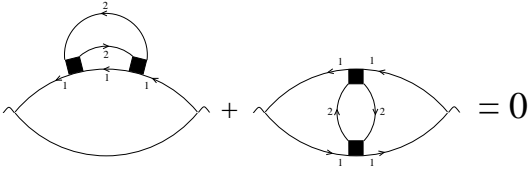}} 
\vspace{0.2cm} 
\caption{Example for the cancellation of the singularity in forward
scattering.  The majority loop can be further dressed in all possible
ways.}
\label{loop} 
\end{figure} 
} 

Before concluding this subsection we have to derive Eq.~(\ref{nu2})
governing the renormalization of the minority mass. Taking the
expression of $F_{12}$ into account from Eq.~(\ref{F12}) we see that
the large ratio $p_{F1}/p_{F2}$ in Eq.~(\ref{m20}) is cancelled in the
angle-dependent term, and hence we obtain a renormalization of the
inverse mass $1/m$ by a quantity of order $1/m_1$.  Utilizing the
expression (\ref{Gamma12}) of $\Gamma_{12}$, we can immediately see,
that further corrections to the first harmonic of $F_{12}$ are smaller
by at least $n_2/N$ and hence result in a nonsingular renormalization
of the mass.  On the other hand, the first harmonic of $F_{22}$ is
nonanalytic, of order $\sqrt{n_2/N}$.  This implies that at zero
minority density the renormalization of the minority mass is due to
the majority ``background'' described by $F_{12}$, and at finite
densities it can be attributed to the residual interaction (described
by $F_{22}$ via the real part of $V$) between minority quasiparticles.
 
To summarize the content of this subsection, we showed by explicit
analysis of all orders of perturbation theory, that the naive
derivation of the Fermi liquid constants for the two fluid model,
Eq.~(\ref{F22ang}), and the minority DOS, Eq.~(\ref{nu2}), is
parametrically justified at $n_2 \ll N$.

\section{Summary}

In conclusion we studied the almost fully polarized two-dimensional
electron liquid. We have shown that assuming (i) the stability of the
majority Fermi-liquid and (ii) the positive renormalization of the
minority mass, no matter how small the density of the minority spin
electrons is, the Fermi two-liquid description is always consistant.
Moreover, microscopic analysis made it possible to find the connection
between different Fermi liquid paramaters, thereby reducing the number
of independent parameters by one.

The established Fermi liquid description enable us to predict
important relations between different thermodynamic observables of the
system, see Subsec.~\ref{experiment}.
  
\section*{Acknowledgments} 
Instructive discussions with A.I. Larkin are gratefully acknowledged.
One of us (I.A.) was supported by the Packard foundation.  Work in
Lancaster University was partially funded by EPSRC and the Royal
Society.  G.Z. thanks the Abdus Salam ICTP for hospitality. We also
thank the Max-Planck-Institut für Physik komplexer Systeme Dresden for
hospitality during the workshop ``Quantum Transport and Correlations
in Mesoscopic Systems and QHE''.

\appendix
\section{}
\label{A} 

In this Appendix we justify the form of the minority self-energy given
in Eq.~(\ref{scalingsigma}) and calculate the dimensionless function
$f(x)$. Clearly, ``ultraviolet'' self-energy diagrams [in the sense
discussed after Eq.~(\ref{scaling})] can be expanded in Taylor series
in both $p^2$ and $\epsilon$. These diagrams are responsible for the
renormalization of the chemical potential (to some negative or zero
value), the minority mass (which is {\em assumed} to be positive) and
the quasiparticle weight $Z_2$. All nonanalytic behavior of the zero
minority density self-energy must come from the ``infrared'', where a
perturbation expansion becomes possible in terms of the small
nonconstant part of the screened interaction.

Our strategy is the following.  We first calculate the contribution to
the imaginary part of the minority self-energy $\Sigma_2(\e, \p)$
coming from the diagram in Fig.~\ref{selfenergy}.  Then, we will argue
that all other infrared diagrams are smaller by a factor of at least
$[p/p_{F1}]^2$.  The contribution of Fig.~\ref{selfenergy}. to the
self-energy is

\beq
\Sigma_2(\e, \p)
= i \int \frac{d^2 q}{(2\pi)^2} \int \frac{d\w}{2\pi} \delta V(\w,\q)
G_2(\e+\w, \p+\q) B_2^2.
\nonumber 
\eeq

\noindent
For the imaginary part we obtain

\beqa
- && \Im \ \Sigma_2(\e, |\p|) = {Z_2}\Im \int \frac{d^2 q}{(2\pi)^2} 
\theta\left(\e-\frac{[\p+\q]^2}{2m_2}\right) \times \\
&& V\left(\frac{[\p+\q]^2}{2m_2}-\e-i0,|\q|\right).
\nonumber
\eeqa

\noindent
The theta function restricts the domain of integration to momenta
$(\p+\q)^2 < 2m_2\e$. Since $\e \ll \e_{F1}$ we can further simplify
the above formula by (i) using the Ward identity Eq.~(\ref{WardB}),
thus effectively setting $B_2=1/Z_2$, and (ii) by expanding the
screened potential $V$ according to Eq.~(\ref{vaprox}) for small
energies and momenta and for $\w \ll v_{F1}q$.  This way we arrive to

\beqa
- \Im \ \Sigma_2(\e, |\p|) && = - \frac{1}{\nu_1 Z_2} 
 \int \frac{d^2q}{(2\pi)^2} \times \\ 
&& \frac{\e - [\p+\q]^2 / 2 m_2}{v_{F1}|\q|} 
\theta\left(2m_2\e - [\p+\q]^2\right). 
\nonumber
\eeqa

\noindent
Calculating the integral and retrieving the real part of the
self-energy in a way to make it analytic on the upper half-plane, we
arrive to Eq.~(\ref{scalingsigma})

\[
- \tilde{\Sigma}_2(\e, |\p|) = 
\frac{\e \sqrt{-2m_2\e+i0}}{p_{F1}} f\left(\frac{p^2}{2 m_2 \e}\right),
\]

\noindent
with the dimensionless function $f$ being

\beq
f(x) = \frac{4}{9\pi} \left\{ 5(1-x)K(x^{1/2})-2(1+4x)E(x^{1/2}) \right\},
\eeq 

\noindent
where $K(y)$ [$E(y)$] denotes the usual complete elliptic integral of
the first [second] kind.

Note, however, that the above calculated expression of the self-energy
does not contain the full real part, since an arbitrary analytic
function of energy (with possible momentum dependence) could still be
added. This means that one cannot retrieve the real part of the
self-energy, an inherently ultravialet quantity, from the
infrared-related imaginary part.

The contribution of higher order infrared diagrams is of order $\max
\{ \epsilon/\e_{F1}, [p/p_{F1}]^2\}$ because of the smallness of the
nonconstant part of the screened interaction $V$ in the infrared
region of momenta and energies.

{ 
\begin{figure}[t]  
\epsfxsize=3 cm  
\centerline{\epsfbox{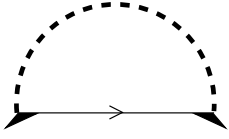}} 
\vspace{0.2cm} 
\caption{Lowest order contribution to the minority self-energy.  The
dashed line denotes the screened Coulomb interaction $\tilde{V}$, see
text, the solid line is the minority propagator $G_2$. Black triangles
will be explained in Subsec.~\ref{3a}.}
\label{selfenergy} 
\end{figure} 
}

\section{}
\label{B}

In Appendix~\ref{B} we outline the derivation of
Eq.~(\ref{gamma22log}), which was obtained using a simple
dimension-counting argument.  As already mentioned, the leading order
contribution to the vertex $\Gamma_{2,2}^{2,2}$ comes from the
irreducible tree level diagrams Fig.~\ref{F22second}. (a), (b), (d)
and (e).
    
Let us first pick diagram (a) and its crossed counterpart, (d) and
write their contribution to the vertex $\Gamma_{2,2}^{2,2}(\P_1, \P_2;
\P_1, \P_2)$ up to an overall factor as

\beqa \label{demo}
\int d^2 q \int d\w \ && G_2(\w, \q) V(\w, \q-\p_1) \times \\  
\left[ 
G_2(\w, \q) \right. && V(\w, \q-\p_2) -
\nonumber
\\
&& \left. G_2(\w, \q+\p_2-\p_1)V(\w, \q-\p_1)
\right].
\nonumber
\eeqa

\noindent
Here the energies of the external legs were put to zero for
simplicity.

In the infrared region ($q<Q$) we deform the frequency integration
contour, expand the renormalized interaction $V$ for $\w<v_{F1}q \ll
p_{F1}$ according to Eq.~(\ref{vaprox}), and introduce the
dimensionless integration variables $x=q/Q, \ y =-2m_2\w/Q^2$.  We
obtain

\beqa
\Gamma_{2,2}^{2,2} = i \gamma_1\left(\frac{\p_1}{Q}, 
\frac{\p_2}{Q} \right)&& \frac{1}{\nu_2} 
\frac{Q^2}{[p_{F1}]^2} \times \\
\int_0^1 xdx && \int_0^{x p_{F1}/Q} dy \frac{y}{(y+x^2)^2},    
\nonumber
\eeqa

\noindent
where the smooth, dimensionless real function $\gamma_1$ describes the
dependence on the external momenta.  This integral is logarithmicaly
divergent at the upper limit of the integration over $y$.  Corrections
may come from the further expansion of the potential $V$: these,
however, result in a convergent integral, and are therefore small
compared to the leading logarithmic behavior.

In the ultraviolet region ($q>Q$) the integration momentum is always
bigger than the external momenta, so we can Taylor expand the integral
around $p_1=p_2=0$. Since the square bracket in Eq.~(\ref{demo})
vanishes for zero external momenta, and because of rotational
invariance, the expansion must start with
$Q^2\gamma_2\left(\frac{\p_1}{Q},\frac{\p_2}{Q}\right)$, $\gamma_2$
being another real dimensionless function. Now we can deform the
contour of the frequency integration and expand $V$ again.
Introducing the dimensionless parameters $x$ and $y$ we obtain

\beqa
\Gamma_{2,2}^{2,2} = && i \gamma_2\left(\frac{\p_1}{Q},
\frac{\p_2}{Q}\right)  \frac{1}{\nu_2} 
\frac{Q^2}{[p_{F1}]^2} \times \\
\int_1^{\infty} xdx && \int_0^{\frac{xp_{F1}}{Q}} dy 
\left(\partial^2_x + \frac{1}{x} \partial_x \right)
\frac{y}{(y+x^2)^2},
\nonumber    
\eeqa     

\noindent
an integral also logaritmically divergent at the upper limit.  

The other two diagrams Fig.~\ref{F22second}. (b) and (e) give similar
contributions. The dressing of the scalar vertices ensures the
cancellation of the quasiparticle weight $Z_2$ of the minority Green
function (omitted in the above estimates for clarity).  Consequently,
one obtains Eq.~(\ref{gamma22log}).

\end{document}